\documentclass[12pt,preprint]{emulateapj}
\begin{document}

\parindent=1.0cm

\title{HAFFNER 16 REDUX: RE-VISITING A YOUNG CLUSTER IN THE OUTER GALAXY 
\altaffilmark{1} \altaffilmark{2}}

\author{T. J. Davidge}

\affil{Dominion Astrophysical Observatory,
\\National Research Council of Canada, 5071 West Saanich Road,
\\Victoria, BC Canada V9E 2E7\\tim.davidge@nrc.ca}

\altaffiltext{1}{Based on observations obtained at the Gemini Observatory, which is
operated by the Association of Universities for Research in Astronomy, Inc., under a
cooperative agreement with the NSF on behalf of the Gemini partnership: the National
Science Foundation (United States), the National Research Council (Canada), CONICYT
(Chile), Minist\'{e}rio da Ci\^{e}ncia,
Tecnologia e Inova\c{c}\~{a}o (Brazil) and Ministerio de Ciencia, Tecnolog\'{i}a e
Innovaci\'{o}n Productiva (Argentina).}

\altaffiltext{2}{This research has made use of the NASA/IPAC Infrared Science Archive,
which is operated by the Jet Propulsion Laboratory, California Institute of Technology,
under contract with the National Aeronautics and Space Administration.}

\begin{abstract}

	Images and spectra recorded with the Gemini Multi-Object Spectrograph on 
Gemini South are used to investigate the stellar content of the 
open cluster Haffner 16. The $(i', g'-i')$ color-magnitude diagram (CMD) 
constructed from these data extends over 10 magnitudes in $i'$, sampling 
the cluster main sequence (MS) and 5 magnitudes of the pre-MS (PMS). 
The fraction of unresolved equal mass binaries 
among PMS stars is estimated to be $0.6 \pm 0.1$. The isochrones do not track the 
PMS on the CMD, in the sense that the PMS has a shallower slope on 
the CMD than predicted by the models. Still, a dip in star counts 
that is associated with the relaxation of PMS stars onto the MS is identified near 
$i' = 17$. The depth and brightness of this feature -- as well as the morphology 
of the cluster MS on the CMD -- are matched by models with a slightly sub-solar 
metallicity that have an age $\sim 20$ Myr and a distance modulus of $12.3 \pm 0.2$. 
A light profile of Haffner 16 is constructed in the W1 filter ($\lambda_{cen} = 
3.4\mu$m) which suggests that the cluster is surrounded by a diffuse stellar halo. 
Spectra are presented of candidate cluster MS and PMS stars 
selected according to location on the CMD. The spectra show characteristics 
that are suggestive of a sub-solar metallicity. H$\alpha$ emission 
is common among objects on the PMS locus on the CMD near $i' = 18$. 
It is suggested that the location of the Haffner 16 PMS on the CMD is affected 
by large-scale cool spot activity, likely induced by rapid stellar rotation.

\end{abstract}


\section{INTRODUCTION}

	Star clusters play an important role as basic astrophysical calibrators. 
It is then ironic that the clusters that are seen today are likely 
the survivors of a process (or processes) that disrupt(s) the vast majority of stellar 
groupings early in their lifetimes (e.g. Lada \& Lada 2003); hence, the clusters 
that provide constraints on -- say -- stellar properties may not be typical of the 
environments where the majority of stars in the Galaxy formed. 
A likely cause of early cluster disruption is 
the rapid outflow of gas driven by massive stars, which 
changes the cluster potential (e.g. Matzner \& Jumper 2015). 
For those few clusters that survive the early loss of mass, the pace 
of subsequent dynamical evolution may depend on factors such as environment 
(e.g. Kruijssen et al. 2012; Silva-Villa et al. 2014), galaxy morphology (e.g. de 
Grijs et al. 2013), and the properties of the molecular clouds from which they form 
(Fujii \& Portegies Zwart 2016). Still, despite the diverse range of potential  
factors that might play a role in cluster disruption, 
a census of star clusters that spans a range of ages and 
masses in nearby galaxies reveals that cohorts of coeval clusters 
are whittled down uniformly by a factor of $\sim 6$ in total number per decade 
in age (Fall \& Chandar 2012). 

	The timescale for the evolution of massive stars is a few Myr. 
Clusters with ages that are less than a few tens of Myr are thus of interest 
for studies of cluster evolution, as they sample a phase when systems 
that survived gas removal are in the early stages of any subsequent evolution. 
Comparing the properties of many such systems may then provide clues 
into the properties of clusters that survive early disruption. 

	Haffner 16 is a young cluster in the outer Galaxy that has not been 
extensively studied in the past. Vogt \& Moffat (1972) present a CMD with a main 
sequence (MS) that is populated by OB stars, thereby pointing to a young age. McSwain 
\& Gies (2005) included Haffner 16 in an investigation of bright emission line sources 
in young clusters. They construct a CMD that goes as faint as $y \sim 17.5$ and assign 
an age of 12 Myr. This age was estimated from stars that are distributed over a 
much wider field than the central regions of the cluster, and so it may be biased 
by non-cluster members. A single Be star candidate was found out of a 
sample of 19 B stars studied.

	Davidge et al. (2013) used deep adaptive optics (AO)-corrected near-infrared 
(NIR) images to examine the stellar content of Haffner 16. An age $\leq 10$ Myr 
was found from the $K$ magnitude of the MS cut-off (MSCO), which is the 
point on the CMD below which stars are still evolving on the pre-MS (PMS). 
A distance modulus of 13.5 was assumed. Davidge et al. (2013) also discuss 
photometry obtained from narrow-band images that suggests 
that many stars with $K \geq 15$ in Haffner 16 have Br$\gamma$ in emission, 
suggesting that chromospheric activity may be common among low mass stars 
in the cluster.

	Haffner 16 was selected for investigation by Davidge et al. (2013) 
in part because existing data suggested that it had a compact size that would 
fit within the $\sim 85$ arcsec science field of the multi-conjugate AO system used 
in that study. However, Davidge et al. (2013) found that PMS stars in Haffner 16 are 
distributed over a larger area on the sky than the brightest cluster members, 
raising the possibility that Haffner 16 might be larger than originally thought. 
Deep photometric and spectroscopic observations of stars in and around 
Haffner 16 that cover many arcminutes on the sky
may thus provide additional insights into its basic properties. 
Wide field observations are also of interest because
stars that are no longer bound to a cluster may linger near it for some time 
(Pfalzner et al. 2014). This raises the possibility of being able to reconstruct the 
primordial mass function (i.e. the mass function prior to disruption) 
of a cluster even if there has been significant dynamical evolution 
and the cluster is in the process of dissolving.

	In the present paper, deep imaging and spectra 
of Haffner 16 that were recorded with the Gemini Multi-Object Spectrograph 
(GMOS) on Gemini South are discussed. The CMD and luminosity function (LF) 
constructed from the images sample sources with masses as low as a few 
tenths solar. In addition to measuring age and reddening, the CMD was also used 
to select objects for follow-up spectroscopy. Multi-slit spectra were obtained of 
92 sources with a common spectral coverage of $0.53 - 0.84\mu$m, although the 
spectra of individual stars extend beyond these wavelength limits. 

	The paper is structured as follows. The imaging and spectroscopic 
observations are described in Section 2, as are the procedures that are 
used to remove instrumental and atmospheric signatures from the raw data. The 
photometric measurements are the subject of Section 3, while 
the cluster light profile constructed from infrared images is presented 
in Section 4. The light profile is used to identify areas on the sky that 
are likely dominated by cluster stars, which is an important issue for Haffner 16 
given the high density of field stars. The CMD and LF are examined in Sections 
5 and 6, while spectra of individual stars are discussed in Section 7. The paper 
closes in Section 8 with a summary and discussion of the results.

\section{OBSERVATIONS \& REDUCTIONS}

	The images and spectra that are the basis of this study were recorded with 
GMOS (Hook et al. 2004) on Gemini South as part of 
program GS-2014A-Q-84 (PI: Davidge). GMOS is the facility visible-light imager and 
spectrograph. The detector was\footnote[1]{The CCDs that make up the GMOS detector 
have since been replaced.} a mosaic of three $2048 \times 4068$ EEV CCDs. Each 
$13.5\mu$m square pixel subtended 0.073 arcsec on the sky. The three CCDs covered 
an area that is larger than that illuminated by the sky as spectra 
may be dispersed outside of the sky field. The images and spectra were both 
recorded with $2 \times 2$ pixel binning. 

\subsection{GMOS Images}

	$g'$ and $i'$ images of Haffner 16 were recorded on the 
night of December 31, 2013. Exposure times and image quality measurements 
are summarized in Table 1. Long and short exposures were recorded 
in both filters so that a CMD that samples stars over a large range of brightnesses 
could be constructed.

\begin{deluxetable}{cll}
\tablecaption{GMOS Images}
\startdata
\tableline\tableline
Filter & Exposures & FWHM \\
 & (sec) & (arcsec) \\
\tableline
$g'$ & $1 \times 1$ & 0.55 \\
 & $1 \times 100$ & 0.55 \\
$i'$ & $1 \times 1$ & 0.45 \\
 & $5 \times 20$ & 0.45 \\
\tableline
\enddata
\end{deluxetable}

	A series of calibration frames were also obtained.
Biases were recorded at the end of the night, and these were median-combined to 
construct a final bias frame.  A fringe frame in $i'$ and 
flat field frames constructed from observations of the twilight sky 
were provided by Gemini as part of the calibration package for this program.

	A standard processing flow for CCD mosaic imaging was applied to 
remove instrumental signatures from the images. 
To start, the output from each CCD was multiplied by its gain, and 
this was followed by bias subtraction, the trimming of 
overscan regions, and division by a flat-field frame. 
The fringe frame, scaled to match the exposure time of the images, was 
then subtracted from the $i'$ images.

	The deep $i'$ images were recorded with 
on-sky dithering. These were aligned using stars across 
the science field as reference points and averaged together after processing. 
All other exposures were shifted to match the reference frame defined by the 
final deep $i'$ image, which is shown in Figure 1. The GMOS science 
field is divided into four regions to facilitate the analysis of the photometry and 
spectra (Section 4), and the boundaries are indicated in Figure 1.

\begin{figure*}
\figurenum{1}
\epsscale{0.95}
\plotone{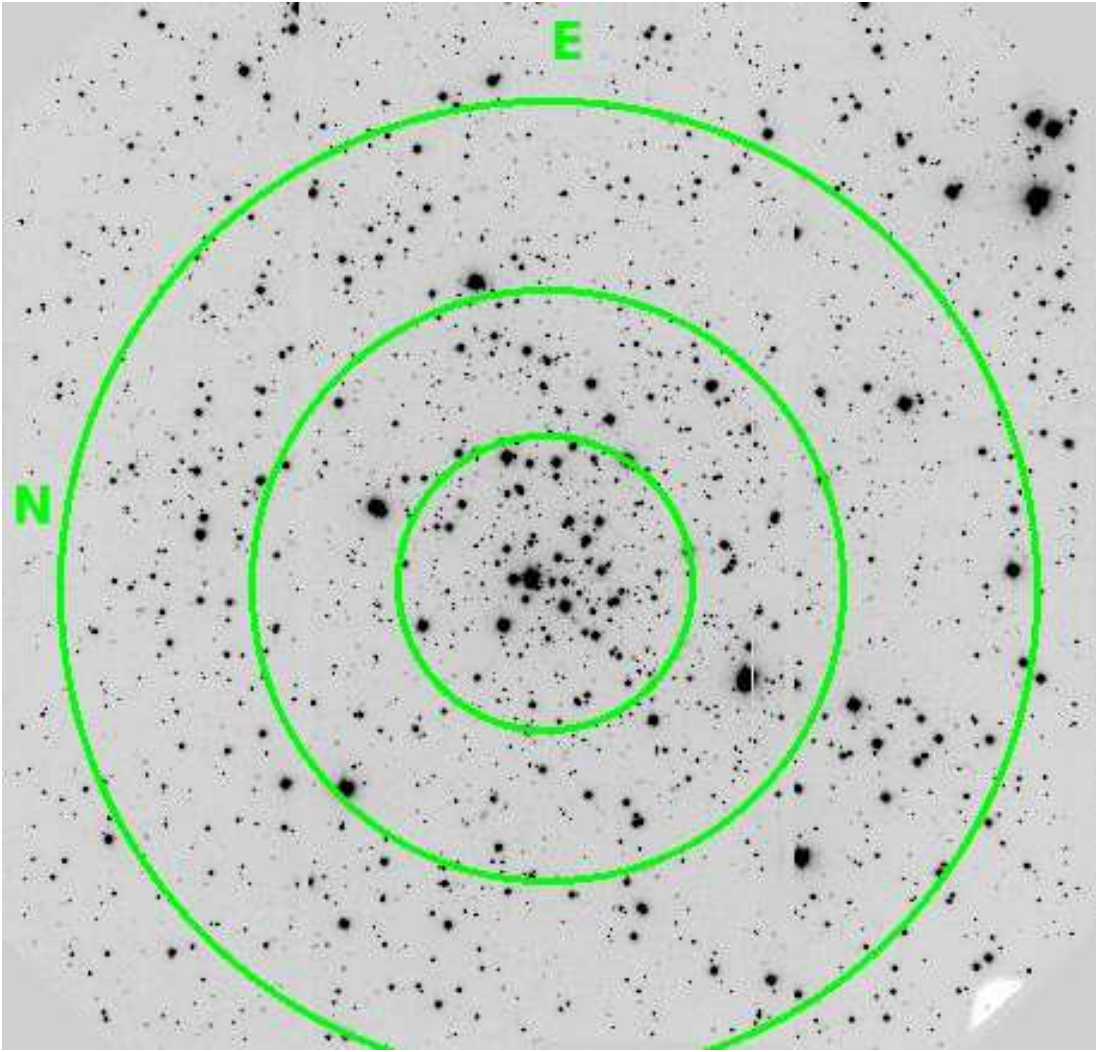}
\caption{Deep $i'$ image of Haffner 16. The image covers $5.9 \times 5.6$ arcmin, with 
North to the left, and East at the top. The green circles mark the boundaries of 
the regions that are used in the analysis of the photometric and spectroscopic 
observations. Working outwards from the cluster center, these are the 
Center, Shoulder, Field 1, and Field 2 regions.}
\end{figure*}

\subsection{GMOS Spectra}

	The spectra were recorded during five nights in March and April 2014. 
The primary criterion for target selection was location on the $(i', g-i)$ CMD. 
High priority was assigned to CMD-selected sources in the middle CCD of 
the mosaic to facilitate common wavelength coverage. 
Masks were designed from images in which the cluster was centered on this CCD 
(Figure 1), and so this detector is also where the density of cluster members is 
highest. Radial velocities measured from these data are of little use for assessing 
cluster membership given the modest resolution of the spectra. 

	Six slit masks were designed, with the target stars in each mask 
having similar magnitudes. The date that each mask was observed, the total 
exposure time, and the FWHM of stars in the slits are listed in Table 2. The 
exposure times reflect the brightnesses of the stars targeted for each mask.

\begin{deluxetable}{clcc}
\tablecaption{GMOS Spectra}
\startdata
\tableline\tableline
Mask & UT Date & Exposure & FWHM \\
\#  & Observed & Time & (arcsec) \\
 & (2014) & (secs) & \\
\tableline
1 & April 3 & $2 \times 150$ & 1.0 \\
2 & March 19 & $2 \times 150$ & 0.9 \\
3 & March 30 & $2 \times 600$ & 1.0 \\
4 & March 27 & $2 \times 600$ & 0.9 \\
5 & April 3 & $4 \times 600$ & 0.9 \\
6 & April 2 & $4 \times 600$ & 0.9 \\
\tableline
\enddata
\end{deluxetable}

	The spectra were dispersed with the R400 grating ($\lambda_{blaze}=7640\AA\ , 
400$ lines/mm), with a GG455 filter deployed to suppress light from higher orders. 
Each mask was observed with two grating rotations such that the 
central wavelengths for a source at the center of the detector mosaic
is either 6750 or 6800\AA. Having spectra with two wavelength settings for each mask 
allows the gaps between the GMOS CCDs to be filled during processing. 

	With the exception of acquisition stars (see below), 
stars selected for spectroscopy were placed in $1 \times 5$ arcsec slitlets. In the 
vast majority of cases the star was positioned near the center of the slit, as measured 
along the spatial direction. However, if another star could be placed in the 
slit with only modest re-positioning of the slit then the mask design was 
adjusted to allow for this. As these additional stars were selected solely based on 
their proximity to a primary target, they are a  
serendipitous sample of disparate objects (Section 7.4).

	The on-sky alignment of each mask was checked 
using three stars that are centered in $2.4 \times 2.4$ arcsec apertures. While the 
primary purpose of these stars is mask centroiding, spectra of these objects 
were also recorded. The alignment stars were selected based on brightness, 
position in the CMD, and location near the edge of the science field. As these objects 
are placed in a wider aperture than the majority of objects then their spectral 
resolution is defined by the seeing disk. However, there is no discernable difference 
between the absorption line profiles of sources in slitlets and those in the 
acquisition apertures. This is likely because the seeing when the 
spectra were recorded was comparable to the 1 arcsec slitlet width (Table 2). 

	The alignment stars tend to be among the 
brightest objects in the spectroscopic sample. The restrictions on 
brightness, position on the CMD, and location in the field (near the edge of 
the science field) limits the number of available 
alignment stars. Hence, some alignment stars were used in multiple masks, 
with the result that more than one spectrum was recorded of some objects.

	Calibration frames were also recorded. 
A frame that monitors the fixed bias pattern was constructed from bias 
observations that were obtained within a few weeks of the spectra. 
Images that monitor pixel-to-pixel variations in sensitivity and variations 
in throughput along the slitlets (`flats') were recorded for each mask 
midway through the on-sky observing sequence. The light source for the flats was a 
continuum lamp in the Gemini Calibration Unit. Spectra 
of a CuAr emission source (`arcs') were recorded for each 
mask at both wavelength settings. Arcs were typically recorded at the end 
of the night or during the following day.

	The processing of the spectra proceeded as follows. The signal from the three 
CCDs were multiplied by their respective gains. The results were 
bias-subtracted and the overscan was trimmed from each CCD. The bias-subtracted 
frames were averaged together after shifting to adjust for wavelength offsets, with 
emission lines in the arc spectra serving as a reference. 
Signal in the gaps between the detectors was ignored when combining the frames. 
Cosmic-rays were identified using an edge-sensing algorithm, and 
then removed by interpolating the signal from surrounding 
pixels. Bad columns and hot pixels were also suppressed by interpolating over 
the affected areas. The flat-field and arc frames were processed in the same way as 
the science frames.

	Slitlets were extracted from the co-added exposures 
for subsequent processing. The signal in each slitlet was divided by a normalized 
flat-field frame that was constructed for that slitlet. Wavelength calibration was 
then done using bright arc lines as reference points. Each 
wavelength-calibrated two-dimenional spectrum was sky-subtracted 
by taking the mean sky level at both ends of the slit on a row-by-row basis. 
If more than one source was in the slit then the sky level 
was measured at a location along the slit where stellar contamination was lowest.

	Individual stellar spectra were extracted from the sky-subtracted 
slitlets by co-adding the signal within the FWHM of each stellar profile. 
The wavelength response of each spectrum at this stage of processing contains 
contributions from the atmosphere, the telescope, GMOS optics, the 
detector, and the flat-field light source. These contributions 
were removed by dividing each Haffner 16 spectrum by the spectrum of the white dwarf 
LTT 4364, which was observed by GMOS in long-slit mode with the same grating, central 
wavelengths, slit width, and order-sorting filter as the cluster 
spectra. The spectrum of LTT 4364 was processed in the 
same manner as the Haffner 16 spectra, and division by this spectrum normalized the 
cluster spectra to that of LTT 4364. Telluric absorption features were also suppressed. 
Residual variations in wavelength response were then removed by fitting a low order 
continuum function to each extracted spectrum.

\section{STELLAR PHOTOMETRY}

	Photometric measurements were made with the 
point-spread function (PSF)-fitting program ALLSTAR 
(Stetson \& Harris 1988). The star lists, PSFs, and preliminary brightnesses used by 
ALLSTAR were generated by running tasks in the DAOPHOT (Stetson 1987) package. 
A single PSF for each image was constructed by combining the signal from bright, 
isolated, unsaturated stars located across the imaged field. Haffner 16 is 
at low Galactic latitudes, and so there is a pervasive population of faint objects, 
some of which inevitably fall within or just outside of the extraction radius 
of the PSF stars. These contaminants were subtracted out using progressively improved 
versions of the PSF.

	The brightnesses of the vast majority of objects were measured by 
fitting the PSF to the central regions of the stellar profiles. 
However, there is a modest number of bright sources that are saturated near their 
profile centers, even in the short exposure images. PSF-fitting was done 
in the PSF wings of these objects. 

	The photometry was calibrated using observations of standard stars in the 
075944--59550 field. The standard stars were observed with GMOS on 
the night of December 30, 2013 (i.e. the night before 
the images for Haffner 16 were recorded). Magnitudes 
for the standard stars in the SDSS system are listed in 
the Southern Standard Stars for the u'g'r'i'z' System website 
\footnote[2]{http://www-star.fnal.gov}.

\section{THE MIR LIGHT PROFILE}

	There is a high density of field stars near Haffner 16, 
and contamination from these objects have the potential to affect an investigation of 
cluster properties. However, the light profile of Haffner 16 can 
be used to identify radial intervals where the light is dominated by cluster members. 
In this section we discuss the light profile of Haffner 16 constructed from 
wide-field images that were recorded as part of the Wide-Field Infrared Explorer (WISE) 
All-Sky survey (Wright et al. 2010). Images taken with WISE cover wavelengths that are 
less susceptible to dust absorption than those at visible or NIR wavelengths and -- 
at least at the short wavelength range of WISE coverage -- sample light 
that is predominantly photospheric in origin. 

	Processed ALL-Sky survey images were downloaded from the WISE archive 
\footnote[3]{http://wise2.ipac.caltech.edu/docs/release/allsky/}. 
Processed images in this part of the sky are restricted to 
the W1 ($\lambda_{cen} = 3.4\mu$m) and W2 ($\lambda_{cen} = 4.6\mu$m) filters. 
The W1 filter samples wavelengths where the contribution from 
photospheric light is expected to dominate the signal, and so the light profile 
was constructed from observations in this filter. 

	The azimuthal averaging technique described by Davidge et al. (2016) was 
used to multiplex the signal from cluster stars and suppress bright field stars. 
In brief, the image is divided into 24 azimuthal zones centered on the cluster. The 
mean signal at each radius is found within each zone. There are then 24 measurements 
of the surface brightness at a given radius, and the median of these is adopted as 
the cluster surface brightness at that radius. A basic assumption is that 
the cluster light follows circular isophotes, and 
the distribution of bright stars near the center of Haffner 16 in Figure 1 
suggests that this assumption is reasonable. A $26 \times 26$ arcsec 
median filter was applied to the images prior to combination 
to suppress light from bright stars, while retaining the light from the numerous 
unresolved stars that make up the main body of the cluster. 

	The light profile of Haffner 16 in the W1 filter is shown in Figure 2. 
The W1 surface brightness measurements were calibrated using 
zeropoints from Jarrett et al. (2011). Background light, which was measured at 
distances in excess of 5 arc minutes from the cluster center, is the dominant source 
of uncertainty in the profile at large radii. The level of this uncertainty 
is illustrated with the green lines in the figure, which show the 
range in surface brightness measurements that result if the sky level 
is varied by $\pm 1 \sigma$, where the dispersion was estimated from sky 
level measurements made in different sub-sections of the field at large offsets 
from Haffner 16.

\begin{figure*}
\figurenum{2}
\epsscale{1.0}
\plotone{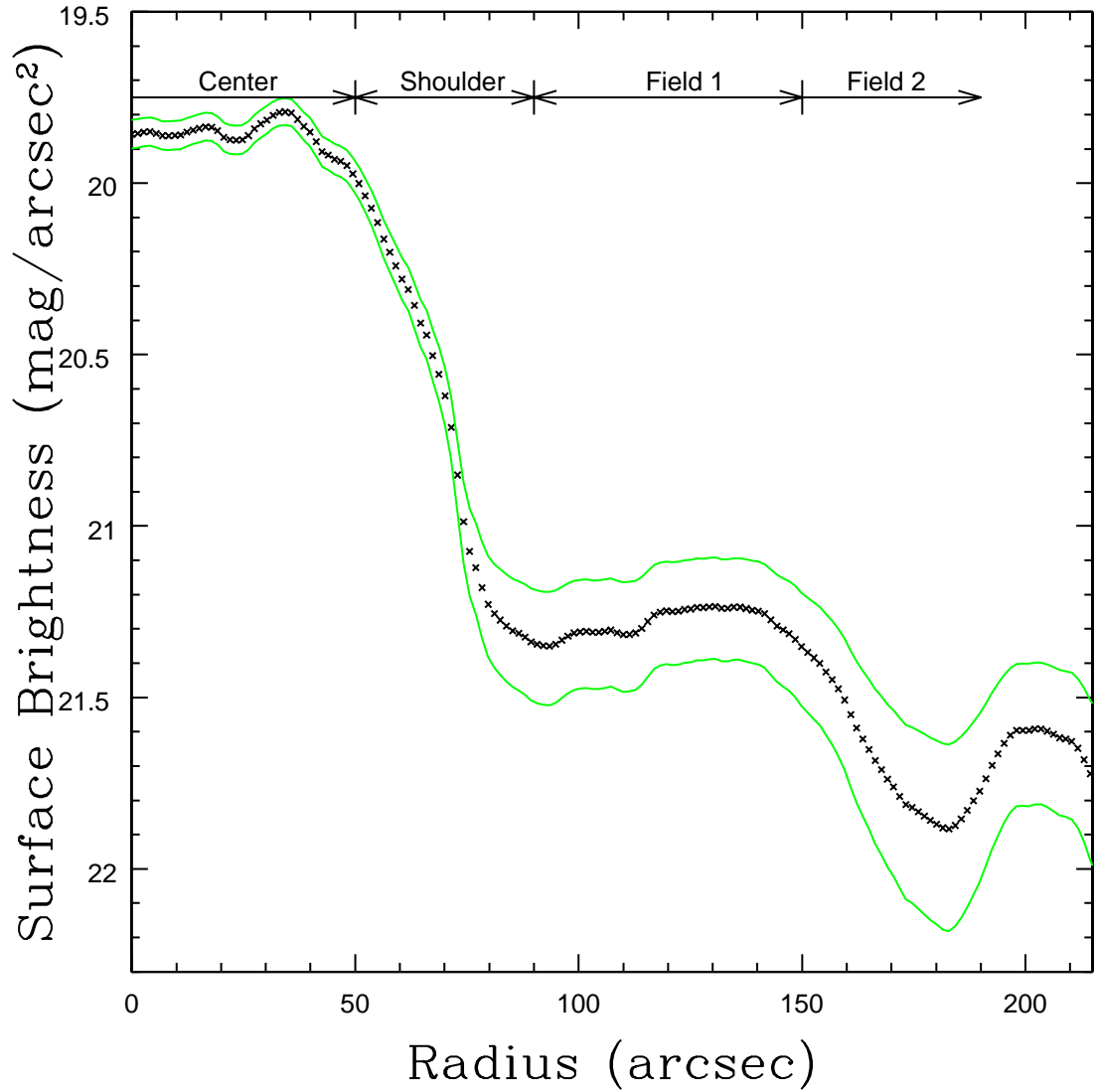}
\caption{Light profile of Haffner 16 in the W1 ($\lambda_{cen} = 3.4\mu$m) filter. 
The profile was constructed using the procedure 
described in the text. The green lines show the surface 
brightnesses that result if the sky level is varied by $\pm 1 \sigma$, 
with the dispersion computed from sky level measurements made at different locations. 
The radial intervals that are used in the analysis of the photometry and spectra are 
indicated.} 
\end{figure*}

	The light distribution in Figure 2 is more-or-less flat within 
50 arcsec of the cluster center, and then drops over a 40 arcsec interval. 
The background-subtracted light level at large radii does not drop 
to zero in the area imaged with GMOS, indicating that there is an over-density of 
light with respect to larger radii throughout the GMOS science field. 
Diffuse stellar halos have been detected around other young clusters (e.g. Davidge 
2012), and some of the stars in the circumcluster area may be present or former 
cluster members, or may have formed at a similar time as the main body of the cluster. 
Indeed, stars that are no longer bound to a cluster may linger near 
it for an extended period of time (Pfalzner et al. 2015), thus forming a halo. 
The collapse of a molecular cloud may involve the formation of both a cluster and 
a diffuse population of surrounding objects (e.g. Bonnell et al. 2011), also resulting 
in a halo.

	For the purpose of the current investigation, the analysis of cluster 
stars is restricted to $\leq 90$ arcsec from the center of Haffner 16. Stars at 
larger radii are deemed to belong to a `field' population. While cluster members 
are almost certainly present at larger radii, the scatter in the CMDs 
presented in Section 5 indicates that non-cluster stars dominate at distances 
in excess of 90 arcsec from the center of Haffner 16.

\section{THE COLOR-MAGNITUDE DIAGRAM}

	The CMDs of stars in the four radial intervals indicated in Figure 2
are shown in Figure 3. The CMDs in the two left hand panels are of regions where 
the integrated light is dominated by cluster stars, while the CMDs in the other 
two panels cover radii where non-cluster stars are 
expected to dominate (Section 4). The points in the CMDs 
with $i' < 17.5$ were obtained from the short exposure images; those 
with $i' < 14$ were obtained by fitting the PSF to the wings of 
stellar profiles to avoid the saturated profile centers. 
All points with $i' > 17.5$ were measured from the long exposure images. 

\begin{figure*}
\figurenum{3}
\epsscale{1.0}
\plotone{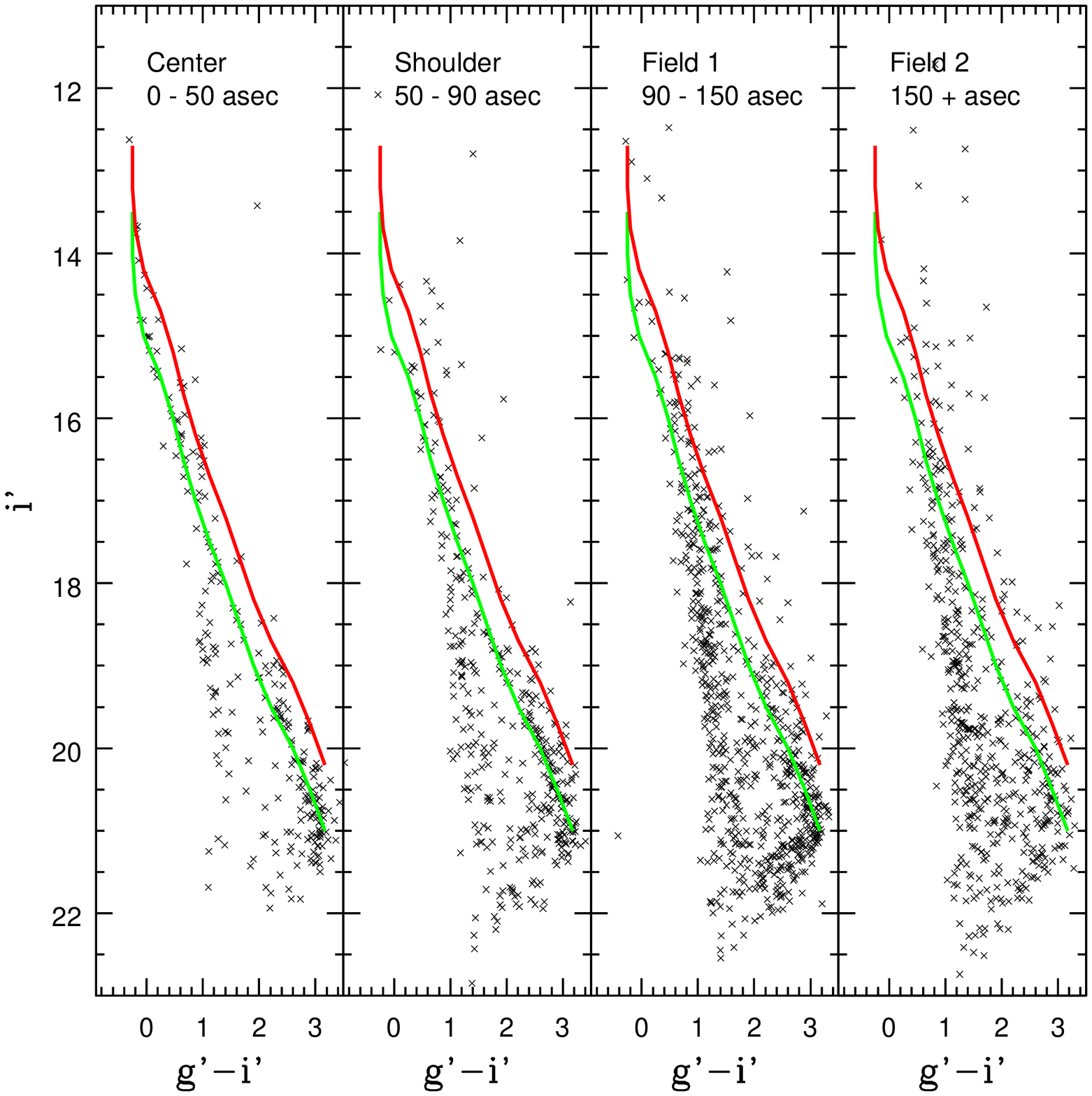}
\caption{$(i', g'-i')$ CMDs of stars in and around Haffner 16. 
The locus of upper MS stars in Haffner 16 is seen in the 
Center region CMD at $i' < 16$, while a well-defined PMS slices diagonally across 
the Center and Shoulder CMD when $i' > 18$. A steep blue sequence that is made up of 
disk stars is seen in the Field 1 and 2 CMDs. The green line is a hand 
drawn representation of the cluster sequence in the Center 
and Shoulder CMDs. The red line is this same fiducial, but 
shifted upwards by 0.75 magnitudes to mark the expected location of 
unresolved equal mass binaries in Haffner 16. There are sources in the Center and 
Shoulder CMDs that fall close to the binary sequence, and in the text it is shown 
that these have markedly higher densities on the sky than objects with similar 
photometric properties in Fields 1 and 2.}
\end{figure*}

	Stars in Haffner 16 define a tight locus of objects in the CMDs of 
the Center and Shoulder regions. A well-defined blue MS is evident at $i' < 16$ in the 
Center region CMD. Cluster MS stars may also be present 
in modest numbers in the CMDs of the Shoulder and Field 1 regions, although the scatter 
near the bright end of those CMDs suggests greater fractional contamination from 
non-cluster stars than at smaller radii. The PMS forms a tight line in the Center and 
Shoulder CMDs that runs from $i' \sim 18$ to $i' \sim 21.5$, where it 
terminates at the faint limit of the CMD near $g'-i' \sim 3$. 
The GMOS images thus sample stars in Haffner 16 that are as faint as $g' \sim 24.5$.

	The green line in each panel is a hand-drawn representation of 
the cluster locus. The upper part of the cluster fiducial is based mainly on the 
Center CMD, whereas at fainter magnitudes it relies on both the Center 
and Shoulder region CMDs. The CMD of Haffner 16 between $i' = 17$ and 18 
is not well populated (Section 6), and so the cluster sequence may be poorly defined at 
those magnitudes. 

	Given the Galactic latitude of Haffner 16, it is not 
surprising that the Field 1 and 2 CMDs are well populated. 
The substantial scatter among the stars with $i' > 15$ in the Field 1 and 
2 CMDs is consistent with the majority of the bright stars in these regions being 
foreground/background objects. As for the faint end, there is a clump of objects in the 
Field 1 and 2 CMDs with colors that are similar to those of PMS stars in Haffner 16. 
The nature of these objects is unclear, and their presence raises concerns 
about the ability to distinguish between faint cluster and non-cluster members.

	While identifying individual cluster stars may be problematic 
(although in Section 7 it is shown that stars with $i'$ between 
18 and 19 that fall on the middle CCD of the GMOS detector mosaic 
tend to have spectroscopic characteristics that are suggestive of cluster 
membership), the spatial distribution and photometric properties of PMS 
stars in Haffner 16 as an ensemble can be characterized statistically 
near the faint end of the CMD. This is demonstrated in Figure 4, where 
the $g'-i'$ color distributions near the faint end of the CMDs are shown. 
Each panel in Figure 4 shows stellar densities in $\pm 0.25$ magnitude 
bins centered on different $i'$ magnitudes. The color distributions that remain after 
subtracting stellar densities in Fields 1 and 2 from the 
densities in the Cluster and Shoulder regions -- and so should 
characterize a cluster population that is free of contamination by non-cluster 
stars -- are shown in the right hand panel.

\begin{figure*}
\figurenum{4}
\epsscale{1.0}
\plotone{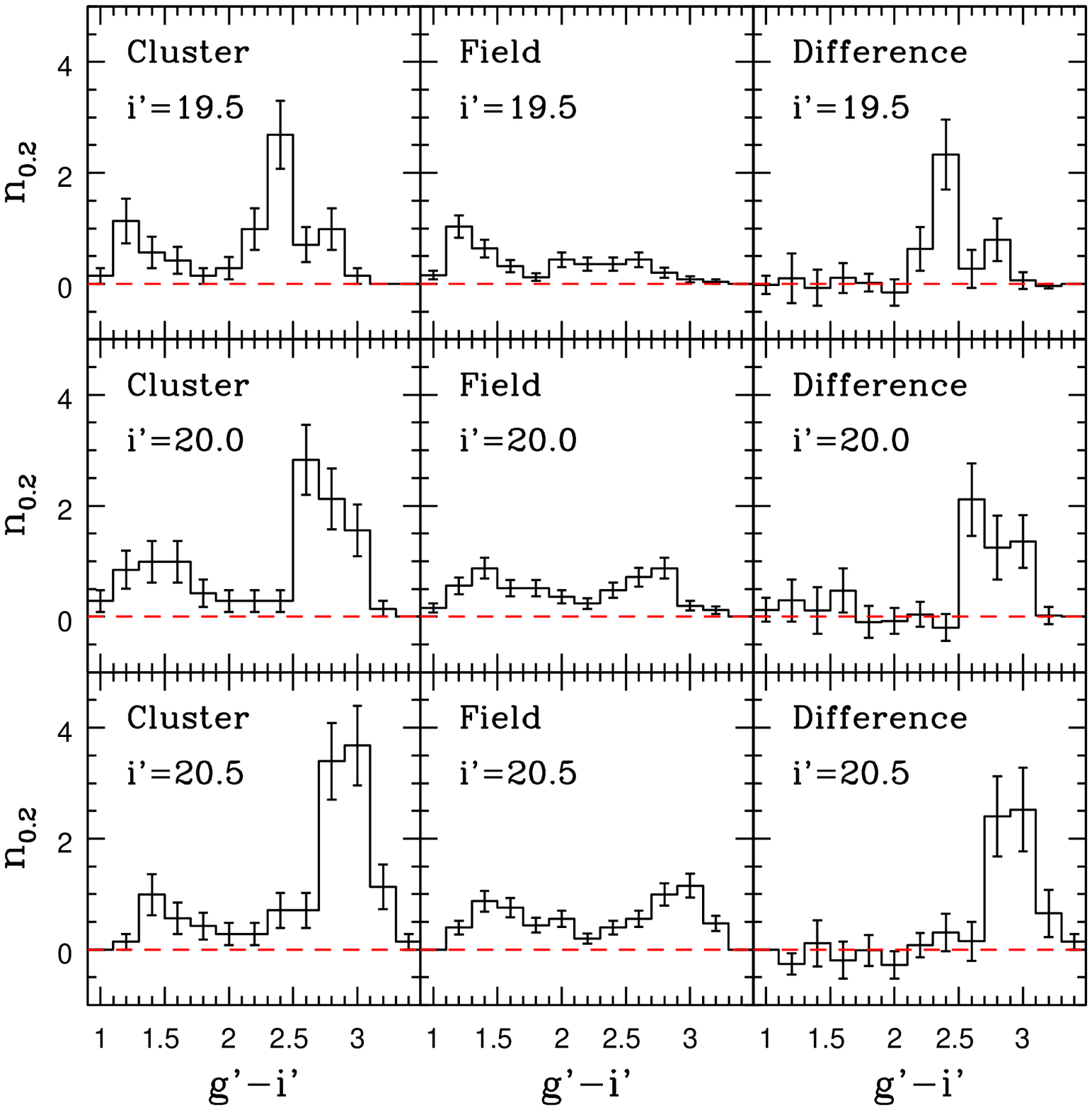}
\caption{$g'-i'$ color distributions near the faint end of the CMDs. 
n$_{0.2}$ is the number of stars arcmin$^{-2}$ per 0.2 magnitude interval in $g'-i'$ 
with $\pm 0.25$ magnitude binning in $i'$. The error bars are $1 \sigma$ 
uncertainties computed with Poisson statistics (i.e. the uncertainty in each bin is 
the square root of the number of counts, with errors in the background counts 
propagated in quadrature). The dashed red line in each panel marks n$_{0.2} = 0$. 
Color distributions are shown for the Cluster$+$Shoulder region (left 
hand column), Field 1 $+$ 2 (middle column), and the difference between the two 
(right hand column). To the extent that Fields 1 and 2 monitor non-cluster members 
then the right hand panel shows the $g'-i'$ distribution of stars in Haffner 
16. Note that objects with $g'-i' < 2.0$ in each magnitude interval 
subtract out, indicating a uniform distribution on the sky -- these 
objects are therefore not cluster members.}
\end{figure*}

	The color distributions in the right hand panel 
contain a population of objects with red $g'-i'$ colors 
that remains at each $i'$ after subtracting the star counts in 
Fields 1 and 2. This residual population is made up of PMS objects in Haffner 16. 
The broad width of the PMS $g'-i'$ distribution is likely the result of a number 
of factors, including binarity (see below), a dispersion in the intrinsic 
photometric properties of the PMS objects, and the $\pm 0.25$ magnitude binning in 
$i'$ that was imposed to obtain statistically significant numbers of sources. Objects 
with $g'-i' < 2$ at each $i'$ in the cluster color distribution are suppressed when 
the Field 1 and 2 counts are subtracted from the cluster counts. This indicates that 
these objects have a uniform distribution on the sky, as expected for stars that are 
not cluster members.

\subsection{The Frequency of Equal Mass Binaries}

	The red line in Figure 3 is the cluster fiducial shifted upwards by 0.75 
magnitudes to mark the expected location in the CMD of unresolved equal mass 
binaries. Processes other than binarity (e.g. star spots) may cause cluster members 
to fall near the binary sequence. However, stars near the binary 
sequence are seen in the Center and Shoulder CMDs over a 
large range of magnitudes, and hence over a range of evolutionary states 
and effective temperatures. This is difficult 
to explain if these sources are not binaries, and so in the remainder 
of the paper these objects are referred to as unresolved equal mass binaries.

	Source counts were made between $i' = 18$ and 20 
in a strip about the red sequence with a width $\pm 0.2$ magnitude in $g'-i'$ 
to determine if unresolved equal mass binaries in Haffner 16 occur in measureable 
numbers. This magnitude range samples the upper regions of the cluster PMS, 
and was selected because it is where the binary sequence is 
well offset from the single star sequence. The densities of objects near 
the red sequence in this part of the CMD are $5.6 \pm 1.6$ arcmin$^{-2}$ (Center), 
$4.7 \pm 1.0$ arcmin$^{-2}$ (Shoulder), and $1.4 \pm 0.2$ arcmin$^{-2}$ (Field 1 
and 2 combined). Thus, there is a statistically significant population of objects 
in the Center and Shoulder regions near the binary sequence that is not present in 
Fields 1 and 2.

	The frequency of unresolved equal mass binaries 
in Haffner 16 can be estimated after measuring the number of stars 
along the single star sequence. The density of sources 
with $i'$ between 18 and 20 that are within $\pm 0.2$ magnitude of the green 
(i.e. single star) fiducial in the Center and Shoulder region is $9.5 \pm 1.2$ 
arcmin$^{-2}$, while in Fields 1 and 2 combined it is $3.2 \pm 0.4$ arcmin$^{-2}$. 
Assuming that the objects near the red sequence in Figure 3 are binaries, then the 
frequency of unresolved equal mass binaries in the Center and Shoulder regions is 
$0.6 \pm 0.1$. This is comparable to the frequency of equal mass binaries among 
MS stars in the open cluster NGC 3105, which Davidge (2017) finds is 
at least 10 Myr older than Haffner 16.

\subsection{Comparisons with Isochrones}

	The combined CMD of the Center and Shoulder regions is compared 
with Z=0.016 PARSEC (Bressan et al. 2012) isochrones in Figures 5 and 6. 
The models were downloaded from the {\it Padova database of stellar evolutionary 
tracks and isochrones} website \footnote[4]{http://pleiadi.pd.astro.it/}.
The composite CMD of Fields 1 and 2 is also shown for comparison. 
Isochrones with other metallicities are considered later in this section.

	Davidge et al. (2013) considered distance moduli of 12.5 and 13.5 for 
Haffner 16, and the results of adopting these are explored in 
Figures 5 and 6. The isochrones have been positioned to match the blue envelope of the 
brightest MS stars in the left hand panel of each figure, and this sets the reddening. 
A distance modulus of 12.5 is adopted for Figure 5. 
The three older isochrones match the inflexion point in the cluster 
CMD near $i' = 15$. The 4 and 10 Myr isochrones fall well above the 
cluster sequence over much of the CMD, while the 32 and 71 Myr 
isochrones fall below the cluster sequence near the faint end of the CMD. The cluster 
sequence is best matched by the 20 Myr isochrone at magnitudes $i' \leq 18$. However, 
at fainter magnitudes the 20 Myr model falls progressively to the left of 
the PMS. There is a $\sim 0.5$ magnitude difference in $g'-i'$ between the 20 Myr 
isochrone and the cluster sequence near $i' = 21$.

\begin{figure*}
\figurenum{5}
\epsscale{1.0}
\plotone{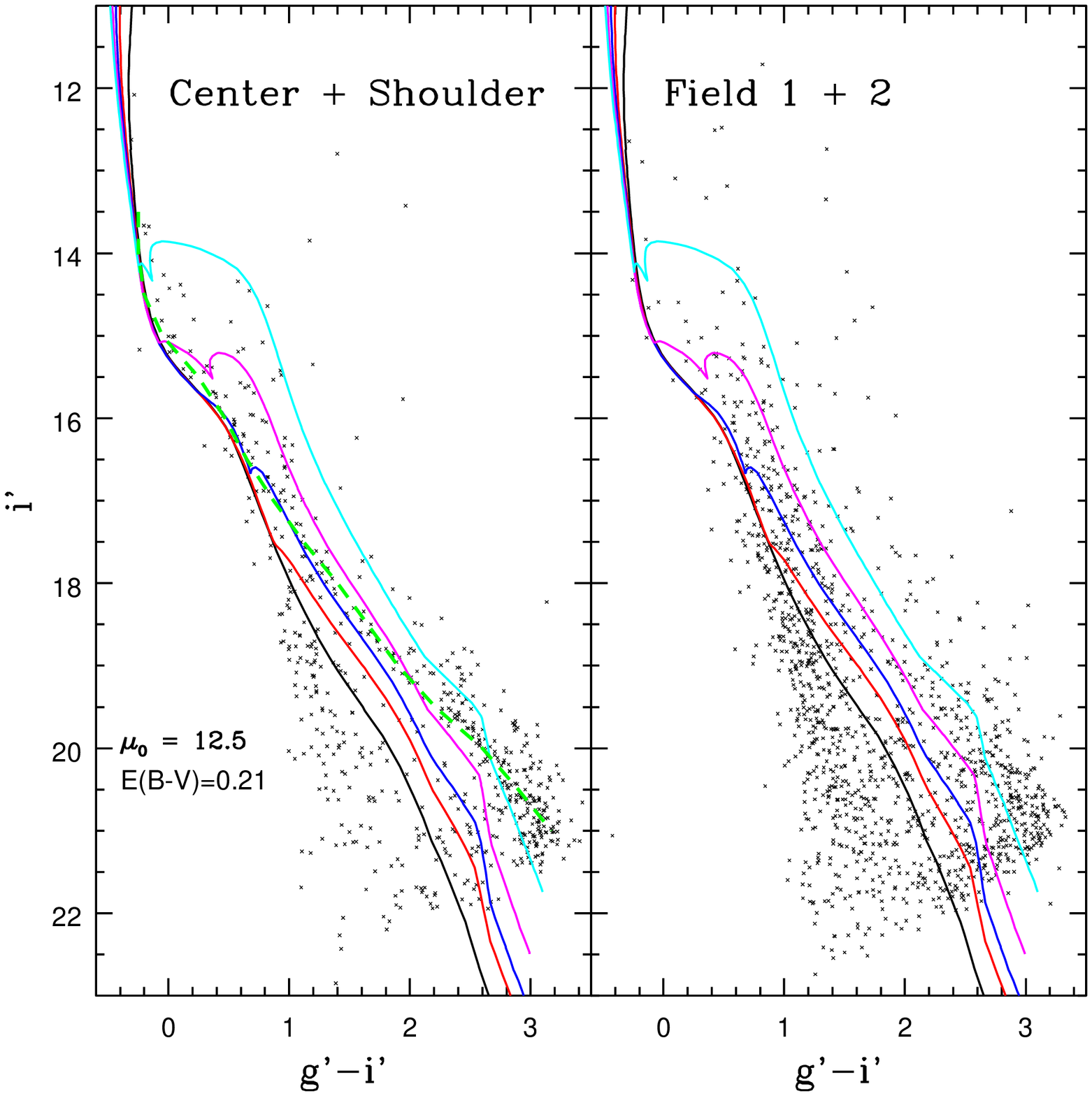}
\caption{Combined CMDs of the Center and Shoulder regions (left hand panel) 
and Fields 1 and 2 (right hand panel) are compared with Z=0.016 isochrones from Bressan 
et al. (2012). A distance modulus of 12.5 is assumed. E(B--V) = 0.21 is found by 
matching the isochrones to the blue edge of the bright MS in the left hand panel. 
E(B--V) was computed from E(g'--i') using the reddening relations in 
Table 6 of Schlegel et al. (1998). Isochrones with ages 4 Myr (cyan), 10 Myr (magenta), 
20 Myr (dark blue), 32 Myr (red), and 71 Myr (black) are shown. 
The dashed green line in the left hand panel is the 
cluster locus from Figure 3. The 4 and 10 Myr models do not match the 
inflexion point in the cluster sequence near $i' = 15$, but come close to 
matching the colors of PMS stars at the faint end. 
While the 20 Myr model matches the cluster sequence when $i' > 18$, it falls 
$\sim 0.5$ magnitudes to the left of the cluster PMS near $i' = 21$.}
\end{figure*}

	A distance modulus of 13.5 is adopted for Figure 6. 
The agreement between the cluster locus and the isochrones is much poorer than in 
Figure 5, and none of the isochrones in this figure match the inflexion point 
in the cluster sequence near $i' = 15$. The isochrones with ages $\geq 20$ Myr pass 
through the blue plume of objects when $i' > 18$, which are shown to be non-cluster 
stars in Figure 4. 

\begin{figure*}
\figurenum{6}
\epsscale{1.0}
\plotone{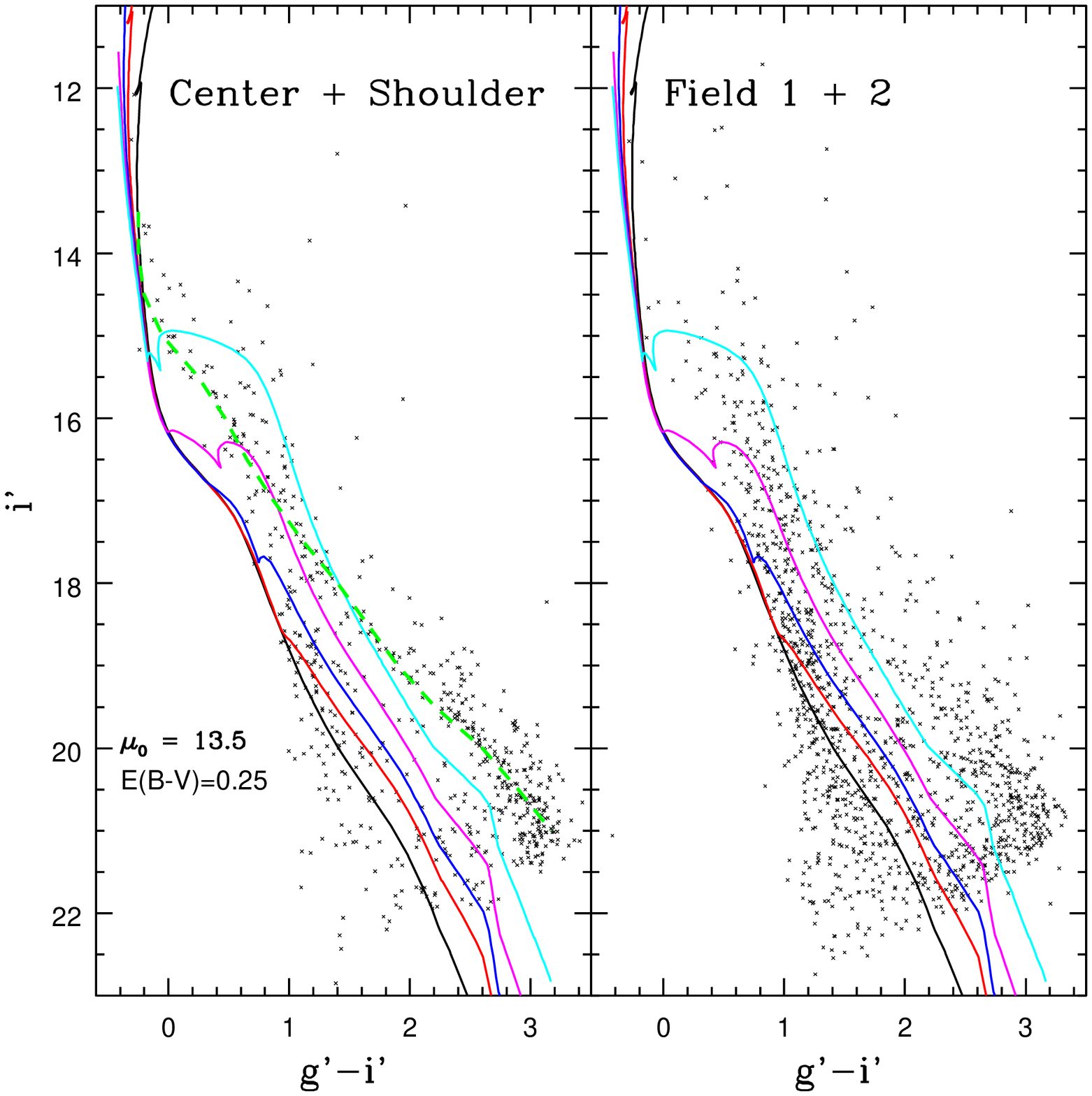}
\caption{Same as Figure 5, but with a distance modulus of 
13.5 and E(B--V) = 0.25. The models with ages 
$\geq 10$ Myr do not reproduce the inflexion point in the CMD near $i' = 15$, 
and all models fall to the left of the PMS at the faint end of the CMD.}
\end{figure*}

	The comparisons in Figures 5 and 6 indicate that a distance modulus close to 
12.5 is favored if Haffner 16 has a metallicity that is near solar. It is also apparent 
that models with a range of ages can match the MS of Haffner 16 at magnitudes 
$i' < 16$. Subtle structure in the CMD that might provide additional insights 
into age and distance is hard to distinguish when the full magnitude range of the 
CMD is shown as in Figures 5 and 6. Therefore, the CMD of Haffner 16 
in a narrower range of magnitudes -- that includes the likely location of the 
MSCO -- is shown in Figure 7. 

\begin{figure*}
\figurenum{7}
\epsscale{1.0}
\plotone{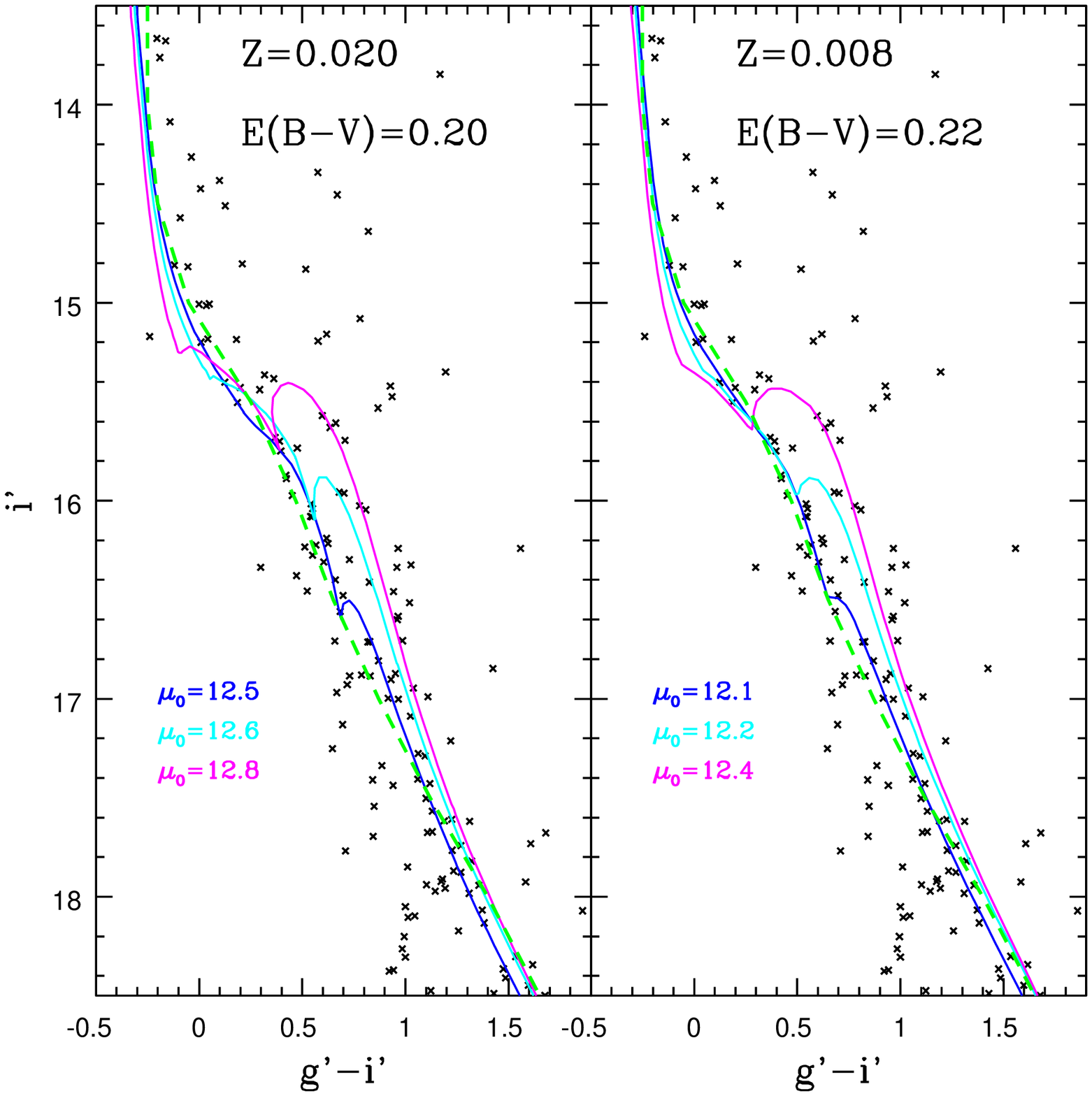}
\caption{Comparing the combined CMD of the Cluster and Shoulder regions 
with Z=0.020 (left hand panel) and Z=0.008 (right hand panel) isochrones near the 
expected brightness of the MSCO. These metallicities span the range seen among 
objects at the Galactocentric radius of Haffner 16. Isochrones from 
Bressan et al. (2012) with ages of 10 (magenta), 14 (cyan), and 20 (blue) Myr are 
shown. The dashed green line is the cluster locus from Figure 3. The 
distance modulus of each isochrone has been adjusted to match the cluster locus. 
The distance moduli found from the Z=0.008 models are 0.4 dex lower than those 
from the Z=0.020 models.}
\end{figure*}

	Haffner 16 might be expected to have a slightly lower than solar metallicity, 
as there is evidence of a metallicity gradient with a slope $\sim -0.04$ dex kpc$^{-1}$ 
among young objects in the Galactic disk (e.g. Daflon \& Cunha 2004; Balser et al. 
2011). If Haffner 16 falls along this trend then it would have a metallicity that 
is $\sim 0.1$ dex lower than solar, and the spectroscopic properties of stars in and 
around Haffner 16 appear to favor a sub-solar metallicity (Section 7).
There is $\sim \pm 0.2$ dex scatter about the mean metallicity trends in the Daflon \& 
Cunha (2004) and Balser et al. (2011) studies. Adopting Z=0.016 as solar, then an upper 
limit for the metallicity of Haffner 16 is Z=0.020, while a lower limit is Z=0.008.
Comparisons are made with models that have these metallicities in Figure 7. 

	The distance modulus for each isochrone has been adjusted to give the 
best `by eye' match to the cluster sequence, and the distance moduli are listed in 
both panels. The distance moduli found with the Z=0.008 isochrones 
are 0.4 dex smaller than those estimated from the Z=0.020 models. 
The two sets of models yield a 0.02 magnitude difference in E(B--V).

	The cluster fiducial is not well-defined 
near $i' = 16$, and this illustrates the difficulty estimating to within a few Myr 
the ages of moderate mass clusters at low Galactic latitudes 
from CMDs alone. For example, the comparisons in Figure 7 indicate that 
if Haffner 16 were as young as 10 Myr then there would be a kink in the 
CMD between $i' = 15$ and 15.5. Small number statistics make the detection 
of such a feature in Haffner 16 problematic. Still, the 10 Myr 
isochrone lies $\sim 0.3$ magnitudes in $g'-i'$ redward of the fiducial 
between $i' = 15.5$ and 17. While there are points with $g'-i' \sim 0.6$ 
and $i' \sim 15.6$ that fall along the 10 Myr isochrone, there is not a 
significant excess density of points in this part of the CMD when compared 
with densities measured in Fields 1 and 2. The stars that fall along the 10 Myr 
isochrone near $i' = 15.6$ also have spectroscopic characteristics that are 
consistent with them being field stars (Section 7). In contrast, 
the stars with bluer colors that fall near the cluster fiducial have spectroscopic 
characteristics that are in line with them being cluster MS stars (Section 7).

	A better match between the 10 Myr isochrone and 
the cluster sequence near $i' = 15$ could be obtained by adopting a distance modulus 
of 12.5 -- 12.6, but then there are difficulties matching the cluster 
locus at fainter magnitudes. In Section 6 it is shown that the $i'$ LF 
of Haffner 16 does not match models with an age of 10 Myr unless a 
distance modulus of 13.8 is adopted, and this is clearly not consistent with the 
CMD in Figure 7. Therefore, the photometric data indicate an age in 
excess of 10 Myr for Haffner 16.

	The 14 Myr sequence is a better match to the observations than 
the 10 Myr model at magnitudes $i' < 16$. Still, the 14 Myr isochrone falls 
to the red of the fiducial at fainter magnitudes. The 14 Myr model has 
a kink in the CMD near $i' = 16$, although the identification 
of this feature in the CMD would likely require more stars than are 
present in Figure 7. Of the three models shown in 
Figure 7, the 20 Myr sequence gives the best overall match to the cluster 
sequence, although the reader is again reminded that there is uncertainty in 
the cluster locus at the magnitudes shown in the Figure. In Section 6 it is shown that 
models with an age of 20 Myr also match the $i'$ LF of Haffner 16. Therefore, based 
on the results in Figure 7 we assign Haffner 16 a distance modulus of 
$12.3 \pm 0.2$.

\section{THE CLUSTER LUMINOSITY FUNCTION}

	The $i'$ LFs of the Center, Shoulder, and combined Center$+$Shoulder regions 
are shown in Figure 8. Only objects detected in both $g'$ and $i'$ have been 
counted. A correction for non-cluster objects was applied by 
subtracting stellar densities measured in Fields 1 and 2. 
As some cluster stars may be present in Fields 
1 and 2, this may result in the over-subtraction of stars, which could 
affect the slope of the LF. However, as the Center and Shoulder regions were selected 
in Section 4 on the basis of having a clear over-density of objects with respect to the 
surroundings then over-subtraction is not expected to have a significant effect 
on the LFs.

\begin{figure*}
\figurenum{8}
\epsscale{1.0}
\plotone{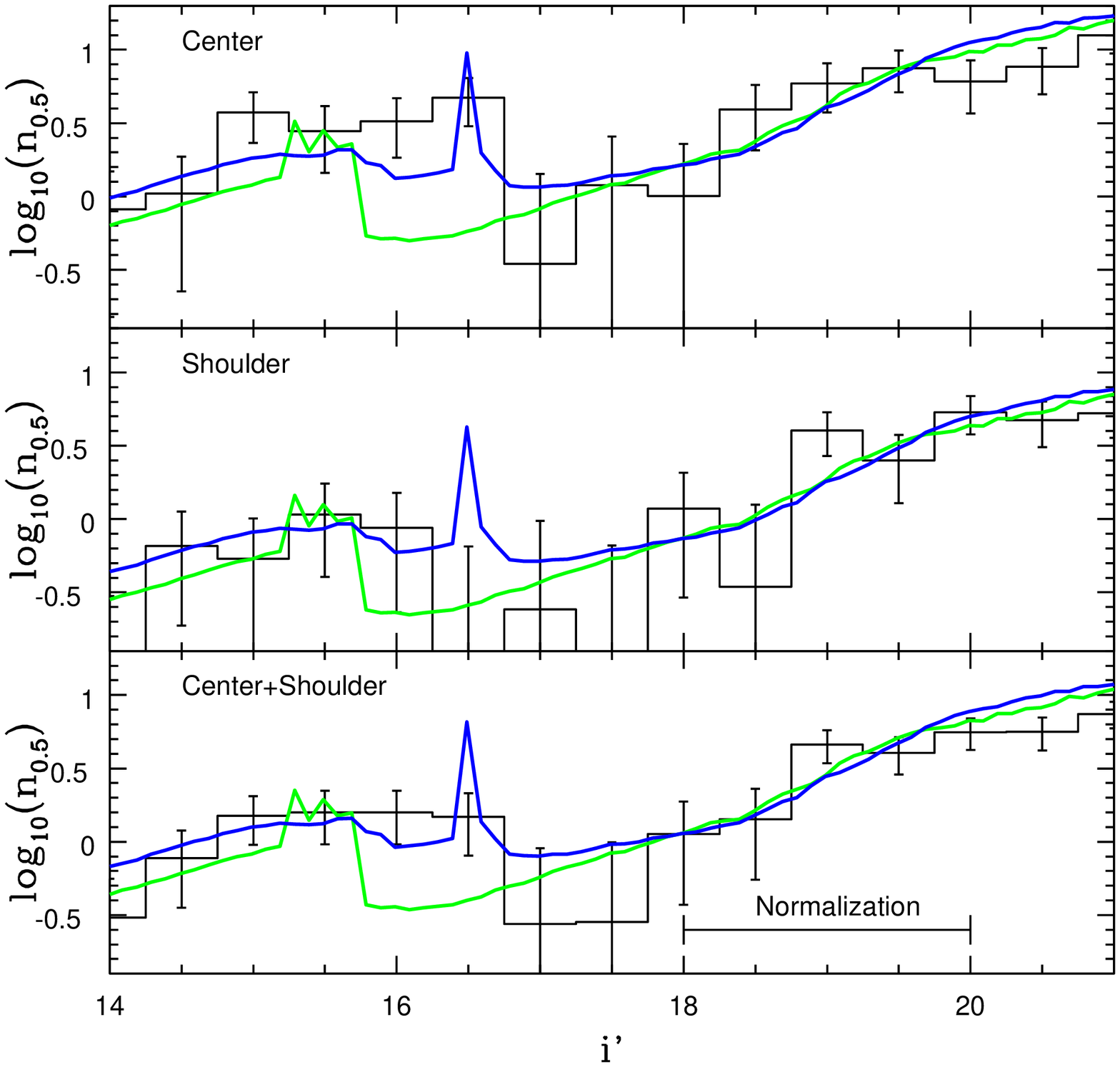}
\caption{$i'$ LFs of the Center, Shoulder, and combined Center$+$Shoulder 
regions. n$_{0.5}$ is the number of objects arcmin$^{-2}$ 
in each 0.5 magnitude interval. The LFs have been 
corrected for non-cluster sources by subtracting number counts from Fields 
1 and 2, scaled to match the areal coverage of the Center and Shoulder 
regions. The error bars are $1\sigma$ uncertainties calculated from Poisson 
statistics. The green and blue lines are model LFs for 
single stars (i.e. no binaries) constructed from Z=0.012 PARSEC isochrones 
with ages of 10 and 20 Myr. The models assume E(B--V) = 0.21 and 
distance moduli of 12.6 (10 Myr) and 12.3 (20 Myr). The models have 
been normalized to match the observed LFs between $i' = 18$ and 20. The 20 Myr model 
matches the Center$+$Shoulder LF within the $\pm 2\sigma$ level at all brightnesses. 
In contrast, the 10 Myr model significantly underestimates the counts at $i' = 16$ 
and 16.5.}
\end{figure*}

	There is a break in counts between $i' =$ 
16.5 and 17.0 in the Center field LF. Although less significant 
statistically, there is also evidence for a similar change in character in the 
LF of the Shoulder region at the same magnitude. The 
$K$ LF of Haffner 16 constructed by Davidge et al. (2013) has a 
similar break, in this case near $K = 16$. The Davidge et al. (2013) 
LF was constructed from a smaller science field, and used counts from the Robin 
et al. (2003) model Galaxy to correct for contamination by non-members. That a 
change in the character of the LF is seen at different wavelengths and fields of view 
indicates that the feature near $i' = 17$ in Figure 8 is not 
due to a fluke over-density of objects in Fields 1 and 2. 
That this feature is seen in both the Center and Shoulder regions also 
indicates that it is not due to a stochastic under-density of objects in one 
region. Rather, the change in the LF near $i' = 17$ (and near $K = 16$ 
in the Davidge et al. LF) probably reflects an intrinsic property of Haffner 16.

	The MSCO can produce a prominent feature in LFs (e.g. 
Cignoni et al. 2010), and the break in the Haffner 16 LF near $i' = 
17$ is likely due to the MSCO. This is demonstrated in Figure 8, where 
model LFs constructed from Z=0.012 (i.e. [M/H] $\sim -0.2$) Padova isochrones 
are shown. As discussed in Section 5.1, Z=0.012 is the metallicity expected for Haffner 
16 if the cluster follows the relations between metallicity and Galactocentric distance 
found by Daflon \& Cunha (2004) and Balser et al. (2011). Based on the 
comparisons in Figure 7, a distance modulus of 12.6 is assumed for the 10 Myr 
model, while a distance modulus of 12.3 is assumed for the 20 Myr model. The models 
in Figure 8 assume a system that is populated by single stars (i.e. no binaries). 
A Chabrier (2003) mass function has been adopted. 

	Model LFs were also generated for Z=0.020 (i.e. [M/H] $\sim +0.1$). 
While not shown here, these models are similar in appearance to the Z=0.012 LFs, 
and provide similar agreement with the Haffner 16 LF after adopting distances based on 
the comparisons between isochrones and CMDs (Section 5). 
This similarity in LF shape indicates that the comparisons in Figure 8 are not 
sensitive to metallicity. 

	The MSCO produces a large discontinuity near $i' = 16$ in the 10 Myr model 
with an amplitude that is similar to the dip in the 
Haffner 16 LF. However, the MSCO in the 10 Myr model occurs 1 
magnitude brighter than this feature in the Haffner 16 LF, and does not match the 
counts near $i' = 16$ and 16.5 in the bottom panel 
of Figure 8. While a distance modulus of 13.6 would produce much better agreement 
between the 10 Myr model and the LF, such a distance modulus is not consistent 
with the morphology of the Haffner 16 CMD, as demonstrated in Figures 6 and 7. 

	The MSCO produces a peak in the 20 Myr model LF near $i' = 16.5$. 
The 20 Myr model matches the LF in the bottom panel of 
Figure 8 at the $\pm 2\sigma$ level from $i' = 14$ to $i' = 20$. 
That the 20 Myr model agrees with the observations within the error 
bars leads us to conclude that the LF is consistent 
with an age $\sim 20$ Myr for Haffner 16. This agreement also indicates that the 
Chabrier (2003) mass function can replicate number counts in Haffner 
16, at least in the mass range probed by these data.

	Davidge et al. (2013) found evidence for mass segregation in Haffner 16, 
in the sense that the brightest MS stars were more centrally concentrated than 
the fainter, lower mass stars in the $K$ LF. However, the LFs of the Center 
and Shoulder regions in Figure 8 are not significantly different, and so there is no 
evidence of mass segregation. The GMOS observations sample a larger area 
than was observed by Davidge et al. (2013) ($5.5 \times 
5.5$ arcmin vs. $1.5 \times 1.5$ arcmin). This allows field star contamination to be 
monitored empirically with the GMOS data, whereas Davidge et al. (2013) 
relied on model star counts. Such models may be uncertain in individual low Galactic 
fields due to -- for example -- localized non-uniformities in the 
line-of-sight extinction.

	In Section 5.1 it was shown that a significant fraction of stars in Haffner 16 
might be in equal mass binaries. Figure 9 shows the Haffner 16 LFs compared 
with models that assume a mix of single stars and equal mass binaries, with a binary 
fraction of 0.6. The models were constructed by combining two single star 
LFs, one of which was shifted to brighter values by 0.75 magnitudes to simulate 
the LF of equal mass binaries. The models in Figure 9 
thus differ from the single star models in that 
there are two MSCO bumps -- the fainter of these is due to single stars, while the 
brighter originates in equal mass binary systems.
These models assume an idealized case in which the component 
stars have not interacted, and are evolving in lockstep.

	The 10 Myr models that include binaries are an inferior 
representation of the observations when compared with the single star models in 
Figure 8. The same is true for the 20 Myr models between $i' = 15$ and 16, although 
at all brightnesses the 20 Myr model in Figure 9 matches 
the observations within the $2\sigma$ uncertainties. 
The binary fraction was measured in a relatively narrow range of magnitudes 
that sampled only the PMS (Section 5.1), and it is possible that it 
may not hold over a broader range of magnitudes. 

\begin{figure*}
\figurenum{9}
\epsscale{1.0}
\plotone{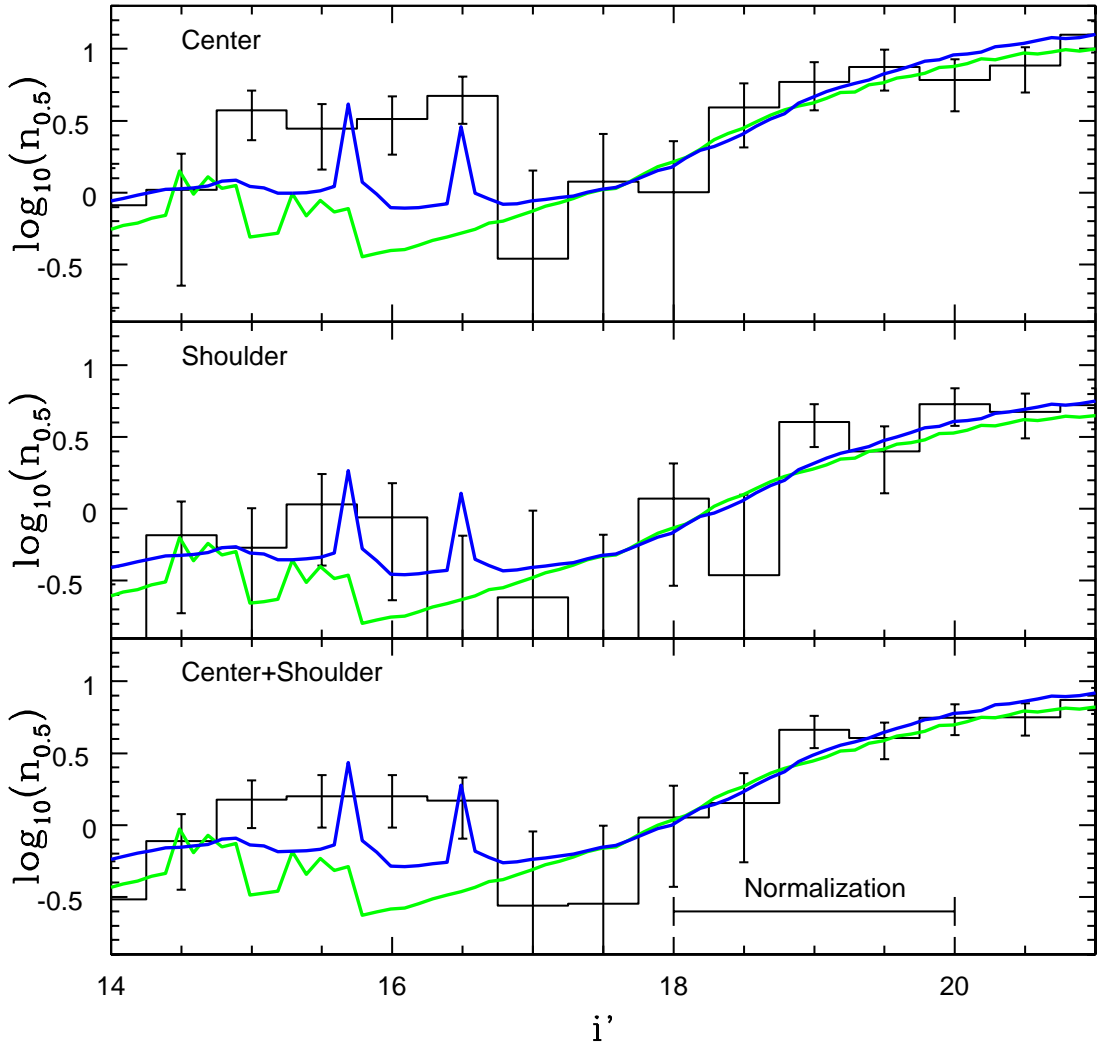}
\caption{Same as Figure 8, but including equal mass binaries. 
The agreement between the models and observations at the bright end 
($i' \leq 16$) is poorer than in Figure 8.}
\end{figure*}

\section{STELLAR SPECTRA}

\subsection{General Properties of the Spectroscopic Sample}

	The $(i', g'-i')$ CMD of the stars targeted for spectroscopy is shown 
in Figure 10. The restricted magnitude range of the targets in each mask is evident. 
The four stars that are in isolated parts of the CMD were 
observed only because they could be placed in the same slit 
as an object that was a primary target for spectroscopy. The spectroscopic 
properties of these objects are discussed in Section 7.4.

\begin{figure*}
\figurenum{10}
\epsscale{1.0}
\plotone{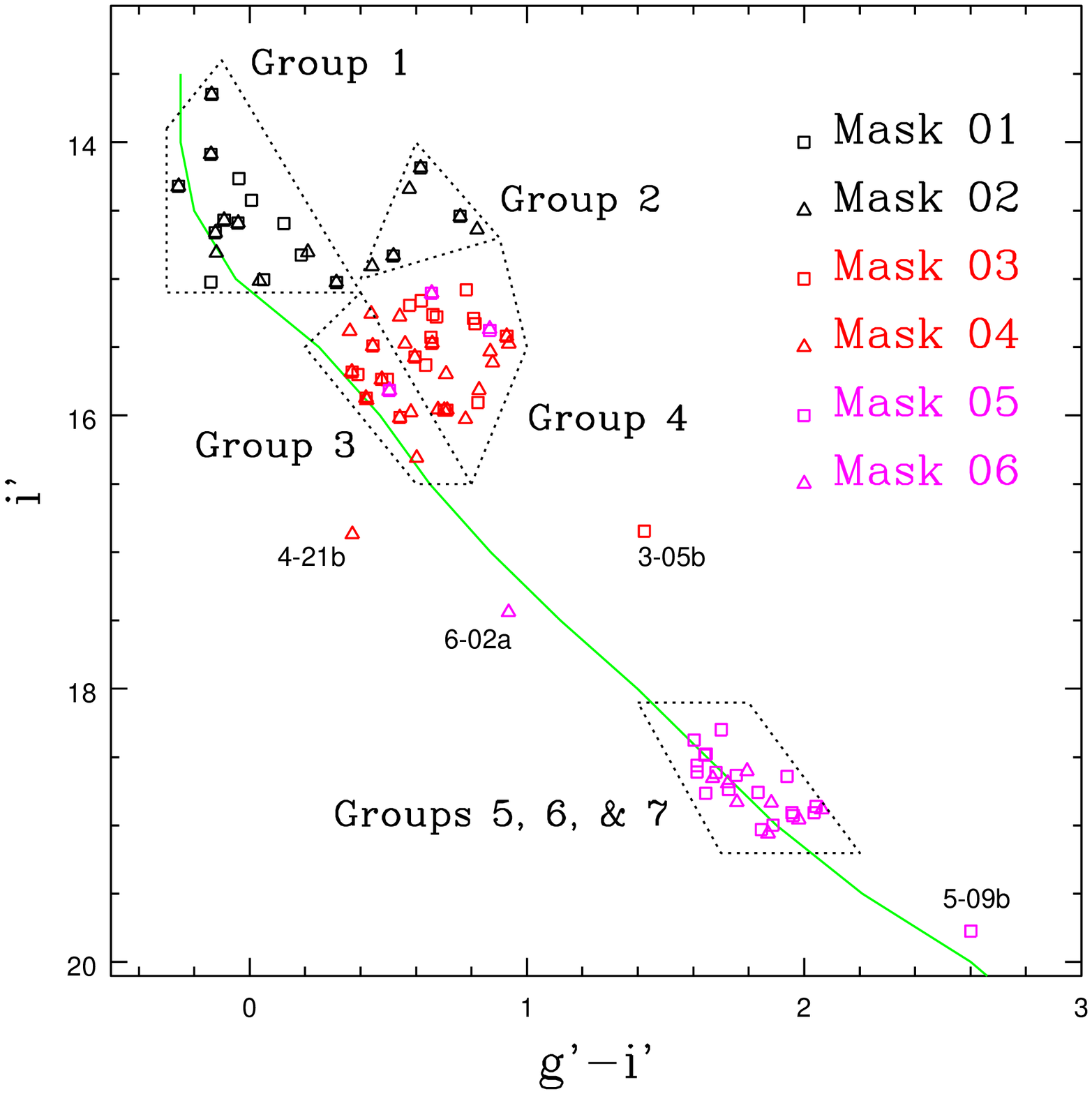}
\caption{CMD of stars with spectra. The green line is the cluster fiducial 
from Figure 3. The areas on the CMD that define the seven groups discussed in 
this paper are indicated. A list of the stars in each group can be found in 
Table 3. The four objects with photometric properties that make them distinct 
from any group are labelled with their identification numbers.}
\end{figure*}

	Haffner 16 is a young cluster, and so it is not surprising that some of its 
members are emission line sources. The location on the CMD of stars that have 
H$\alpha$ either in absorption (black squares) or emission (magenta squares) are 
indicated in Figure 11. None of the stars with $i' < 18$ show obvious H$\alpha$ 
emission, in agreement with the low incidence of emission found among early-type stars 
in this part of the sky by McSwain \& Gies (2005). Weak 
line emission might be hard to detect in the deep H$\alpha$ profiles 
of the stars at brighter magnitudes.

\begin{figure*}
\figurenum{11}
\epsscale{1.0}
\plotone{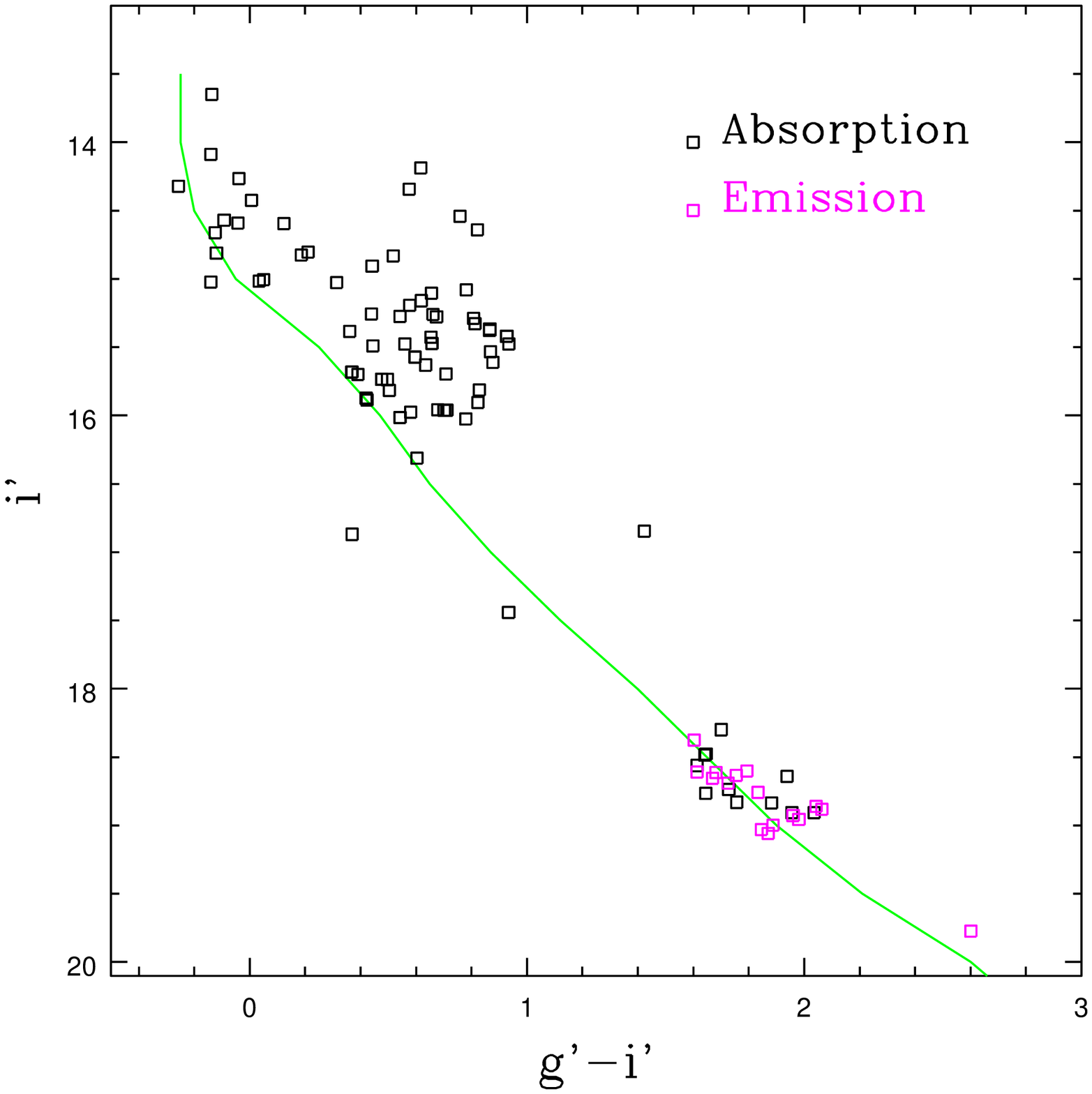}
\caption{CMD of stars with H$\alpha$ in absorption (black squares) 
and emission (magenta squares). H$\alpha$ emission is only detected in stars 
that have $i' > 18$.}
\end{figure*}

	Figure 12 shows the locations of the spectroscopic targets in the 
GMOS science field, with stars marked according to whether or not they have 
H$\alpha$ in emission. The stars that have H$\alpha$ 
in absorption are scattered more-or-less uniformly across the field, whereas the 
emission line sources tend to be concentrated in the Center and Shoulder regions. This 
is due in part to the (intentional) bias to obtain spectra of faint 
stars in the main body of the cluster, where the incidence of cluster members 
is highest. Still, some emission line sources are seen outside of the Shoulder region.

\begin{figure*}
\figurenum{12}
\epsscale{1.0}
\plotone{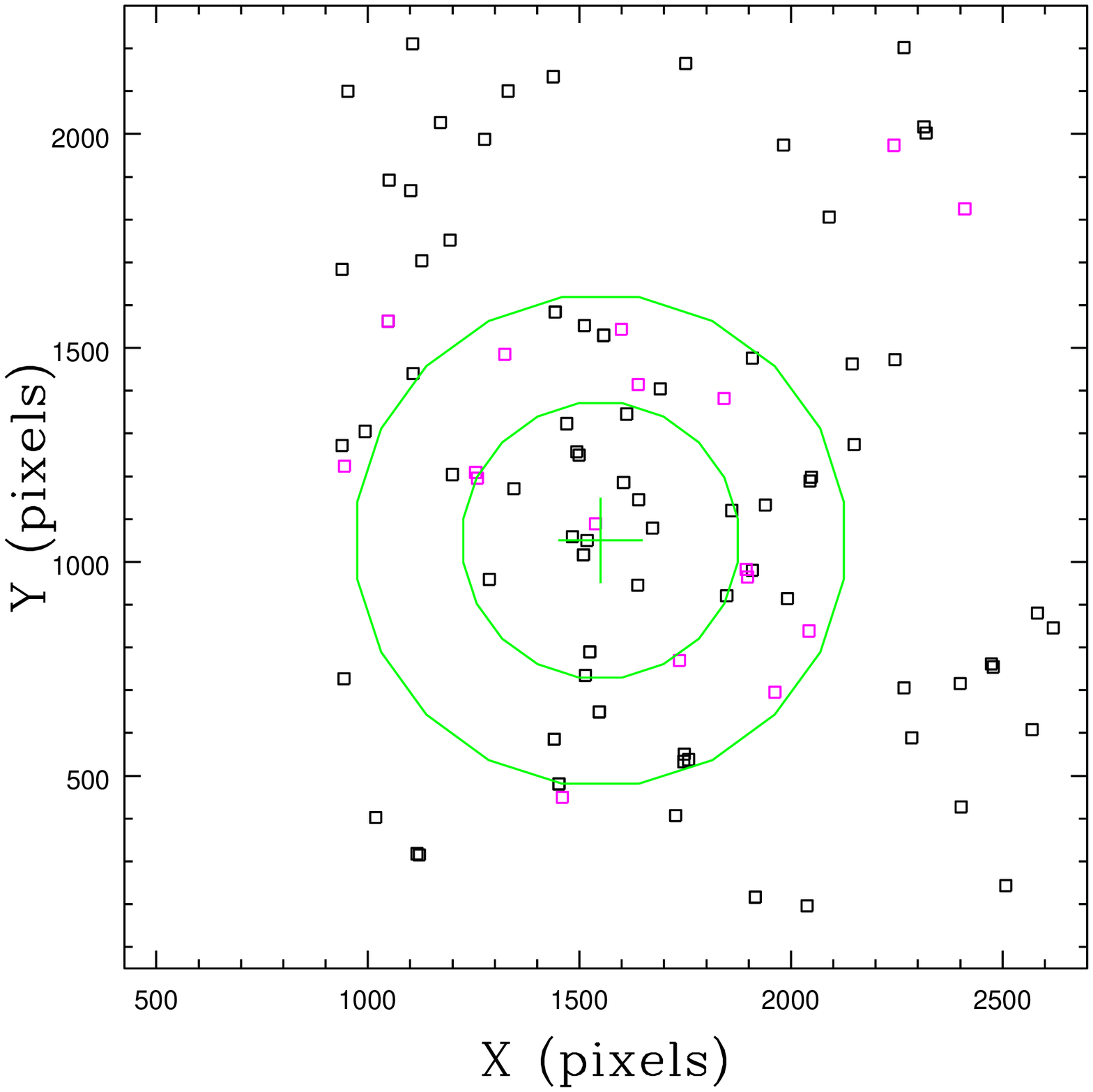}
\caption{Locations of sources that have H$\alpha$ in absorption (black squares) and 
emission (magenta squares). $X$ and $Y$ are pixel co-ordinates on the GMOS detector, 
and these are listed in Table 3 for each source. The area shown is the same 
as that displayed in Figure 1. The center of Haffner 16 
is marked with a cross, and the circles indicate the boundaries of 
the Center and Shoulder regions. The majority of emission line sources fall within 
or just outside of the Shoulder region.}
\end{figure*}

	The stars selected for spectroscopic follow-up have a range of photometric 
properties, and were sorted into seven groups according to 
location on the CMD. The group boundaries are indicated in Figure 10. 
Groups 1, 3, 5, 6, and 7 are made up of objects on or near the cluster sequence. 
Groups 5, 6, and 7 overlap on the CMD, and the behaviour of the H$\alpha$ line 
is used to distinguish between members of these three groups: stars in group 5 
have weak H$\alpha$ absorption, while stars in groups 6 and 7 have either weak (group 
6) or strong (group 7) H$\alpha$ emission.

	The members of each group are listed in Table 3. The identification number 
for each target consists of the mask \# (1 -- 6), followed by 
the slit number, defined such that slitlet \# 1 for each mask is the 
closest to the bottom of the science field. The `a' and `b' 
designations indicate stars that are in the same slitlet. 
The limited number of stars with $i' < 16$ 
causes significant overlap in the object lists for Masks $1 + 2$, and $3 + 4$.
The co-ordinates and photometric properties of stars 
with spectra are given in Table 4. Stars that have 
multiple observations are indicated with an asterix, 
with their identification numbers in other masks given in brackets. 

	A median spectrum was constructed for each group, and the results 
in the 5300 to 6800\AA\ interval, which is the wavelength interval 
that is common to most stars, are shown in Figure 13. 
The spectra in this figure have been convolved with a Gaussian to produce 
a resolution of 640 to improve the S/N ratio of the spectra of stars in groups 5, 6, 
and 7. This spectral resolution is sufficient to allow absorption features that are 
diagnostics of basic stellar properties to be detected (e.g. Worthy et al. 1994). Aside 
from Na D and H$\alpha$, many of the absorption features at these wavelengths 
are Fe I and Ca I transitions, although other atomic and molecular species 
contribute (e.g. Wallace et al. 2011). Applying Equation 9 of Poznanski 
et al. (2012) suggests that interstellar absorption towards Haffner 16 likely 
contributes an equivalent width of $\sim 1\AA$ to the Na D lines.

\begin{deluxetable}{ll}
\tablecaption{Group Membership}
\startdata
\tableline\tableline
Group & Star IDs \\
\tableline
1 & (1-01), (1-02), 1-03, 1-04, (1-05) \\
 & 1-06, 1-07, 1-08, (1-09), (1-10) \\
 & (1-11), 1-12, 1-14, (2-01), (2-02) \\
 & 2-03, (2-05), 2-06, 2-07, (2-09) \\
 & (2-10), (2-11) \\
2 & (1-13), (1-15), (1-16) , 2-04, 2-08 \\
 & 2-12, (2-13), (2-15), (2-16) \\
3 & 3-05a, (3-07), (3-09), (3-16), (3-17) \\
 & 3-21, (3-22), 4-03, (4-07), (4-09) \\
 & 4-12, 4-14a, 4-14b, (4-17), (4-18) \\
 & 4-19, 4-21a, (4-23), (5-15), (6-07) \\
4 & (3-01), 3-02, 3-03, (3-04), (3-06) \\
 & 3-08, 3-10, 3-11, 3-12, 3-13 \\
 & 3-14, 3-15, 3-18, (3-19), 3-20 \\
 & (4-01), 4-02, (4-04), 4-05, (4-06) \\
 & 4-08, 4-10, 4-11a, 4-11b, 4-13a \\
 & 4-13b, 4-15, 4-16, (4-20), (4-22) \\
 & (5-01), (5-19), (6-01), (6-10) \\
5 & 5-02, 5-04, 5-08, 5-14a, 5-16 \\
 & 5-17, 5-18, 5-20, 5-21, 6-02b \\
 & 6-03 \\
6 & 5-03, 5-05, 5-09a, 5-10, 5-11 \\
 & 5-12, 5-13, 5-14b, 6-06 \\
7 & 5-06, 5-07, 6-04a, 6-04b, 6-05 \\
 & 6-08, 6-09 \\  
\tableline
\enddata
\end{deluxetable}

\begin{figure*}
\figurenum{13}
\epsscale{1.0}
\plotone{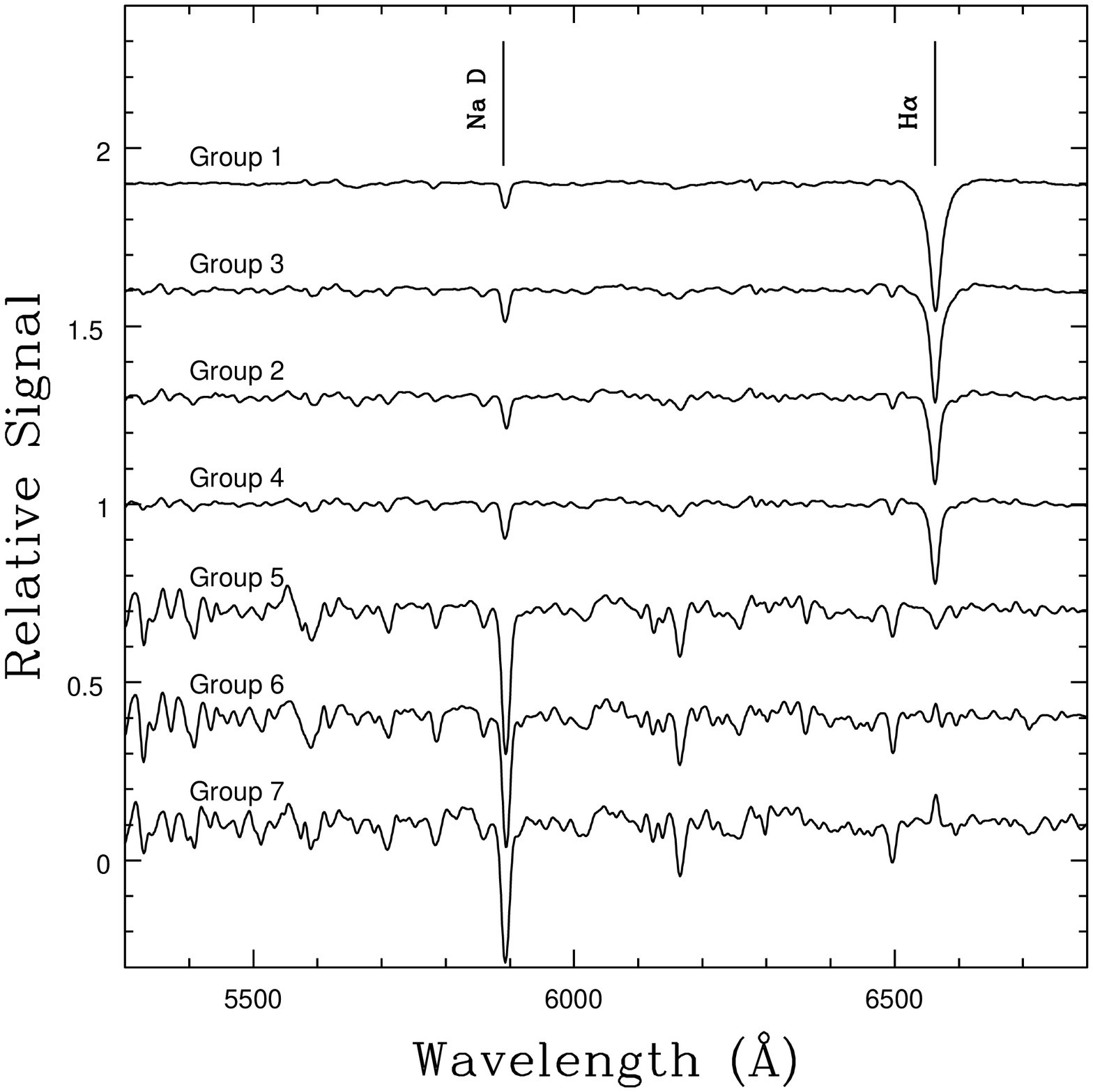}
\caption{Median spectra. The spectra have been shifted vertically 
for display purposes, and have been smoothed to a resolution of 640. Note that 
the spectra are not ordered sequentially by group number.}
\end{figure*}

	Some stars in groups 1 -- 4 were observed in more than one mask, 
and so the S/N ratio of the processed spectra can be estimated in a statistical manner 
by comparing the depths of features in multiple spectra of the same star. 
Groups 1 -- 4 sample stars with $i' < 16$, and 
the spectra in these groups have comparable S/N ratios. 
Consider the depths of Na D and H$\alpha$, which are the strongest features 
in the spectra of stars in these groups. The dispersion 
in the depths of Na D among stars with multiple spectra in groups 1 - 4 is 
only $\pm 1\%$, whereas the dispersion in the depth of 
H$\alpha$ is $\pm 2\%$. The dispersions include contributions from photon 
noise, the centering of the star in the slit, which can affect line shape, 
sky subtraction, and the placement of the continuum.

	There is a star-to-star dispersion in the spectroscopic properties of 
the members of each group. This scatter is due to slight differences among cluster 
members in each group (recall that the groups sample stars with a range of 
magnitude and color) and contamination from non-cluster members. As with the 
examination of the spectra of stars with repeat measurements, the 
star-to-star scatter in the depths of Na D and H$\alpha$ 
provides insights into the range of spectroscopic properties in each group. Groups 
1 and 3 have the smallest star-to-star scatter, with $\sigma = 2\%$ for Na D, 
and $\pm 3\%$ for H$\alpha$. These dispersions are only slightly larger than those 
estimated in the previous paragraph that are due to random and systematic errors 
in individual spectra. The star-to-star scatter is larger for groups 2 and 4, 
with $\sigma = \pm 4\%$ for Na D, and $\pm 6\%$ for H$\alpha$. That 
groups 2 and 4 sample a more diverse range of spectroscopic properties is 
not unexpected, as these groups are made up of stars that 
are offset from the cluster locus on the CMD, and so are likely members of the 
field population. 

	The star-to-star dispersions in the depths of 
Na D and H$\alpha$ among the spectra of stars in groups 5, 6, and 7 are similar, 
with $\sigma = \pm 7\%$ for Na D, and $\pm 5\%$ for H$\alpha$. These 
groups sample fainter stars than those in groups 1 -- 4, and the lower S/N ratio 
of the spectra likely contributes to the larger dispersion in the Na D depths. 
Unlike is the case for groups 1 -- 4, the star-to-star dispersion in H$\alpha$ 
is smaller than that for Na D. This may be due to the
membership of groups 5 -- 7 being based on H$\alpha$ properties, which has the 
potential to bias the H$\alpha$ dispersion measurements within each group.

	A cursory examination of the spectra in Figure 13 reveals 
trends that are broadly consistent with the criteria used to define the groups. 
The group 1 spectrum has the weakest metallic absorption features and deepest 
H$\alpha$ absorption, as expected if the stars in group 1 have earlier 
spectral-types than those in other groups. H$\alpha$ absorption in group 3 is weaker 
than in group 1 and metallic absorption features are more pronounced, which 
is consistent with the stars in group 3 being intrinsically fainter 
than those in group 1. 

	In contrast to groups 1 and 3, the spectra of groups 2 and 4 are very 
similar. The median spectra of these groups have characteristics that are consistent 
with typical member stars having later spectral-types than those in groups 1 and 3, in 
agreement with their $g'-i'$ colors. That the median spectra of groups 2 and 4 do 
not show magnitude-related differences is consistent with the majority of stars in 
these groups belonging to a field population that samples a range of distances. 
The projected distribution of objects in groups 2 and 4 on the sky is 
also such that the majority are located outside of the Shoulder region.

	If Haffner 16 had a very young age then some PMS stars might be 
expected in the parts of the CMD that sample groups 2 and 4. However, 
while falling to the right of the cluster sequence in Figure 10, 
the objects in groups 2 and 4 are likely not dominated by PMS stars in Haffner 16. Even 
adopting an age of 10 Myr and a distance modulus of 12.6 -- which together 
are favorable for the presence of bright PMS stars in Haffner 16 -- then PMS stars 
in Haffner 16 are expected to have $i' > 15$ (Figure 5), placing any such objects 
in group 4, and not in group 2. That the median spectra of groups 2 and 4 are similar 
thus indicates that if PMS stars are present then they are not the dominant population 
in Group 4. 

	Aside from differences in the behaviour of H$\alpha$ that was specified to 
define group membership, the spectra of Groups 5, 6, and 7 are 
similar. The characteristic features of these spectra are 
indicative of a much later spectral-type than stars in the other four groups. The 
spectroscopic properties of the median spectra of the stars in these -- 
and the other -- groups are discussed in greater detail below.

\subsection{Stars with H$\alpha$ in Absorption}

	Composite spectra in the wavelength interval 5300 -- 6800\AA\ 
of the groups that have H$\alpha$ in absorption are shown in 
Figures 14 (groups 1 and 3), 15 (groups 2 and 4), and 16 (group 5). 
Spectra from the Le Bourgne et al. (2003) library, smoothed 
to the same resolution as the Haffner 16 spectra and normalized to the continuum, 
are also shown. The majority of the stars in the Le Bourgne et al. library are 
bright and nearby. Thus, as a group they are expected to have {\it roughly} 
solar metallicities. 

\begin{figure*}
\figurenum{14}
\epsscale{1.0}
\plotone{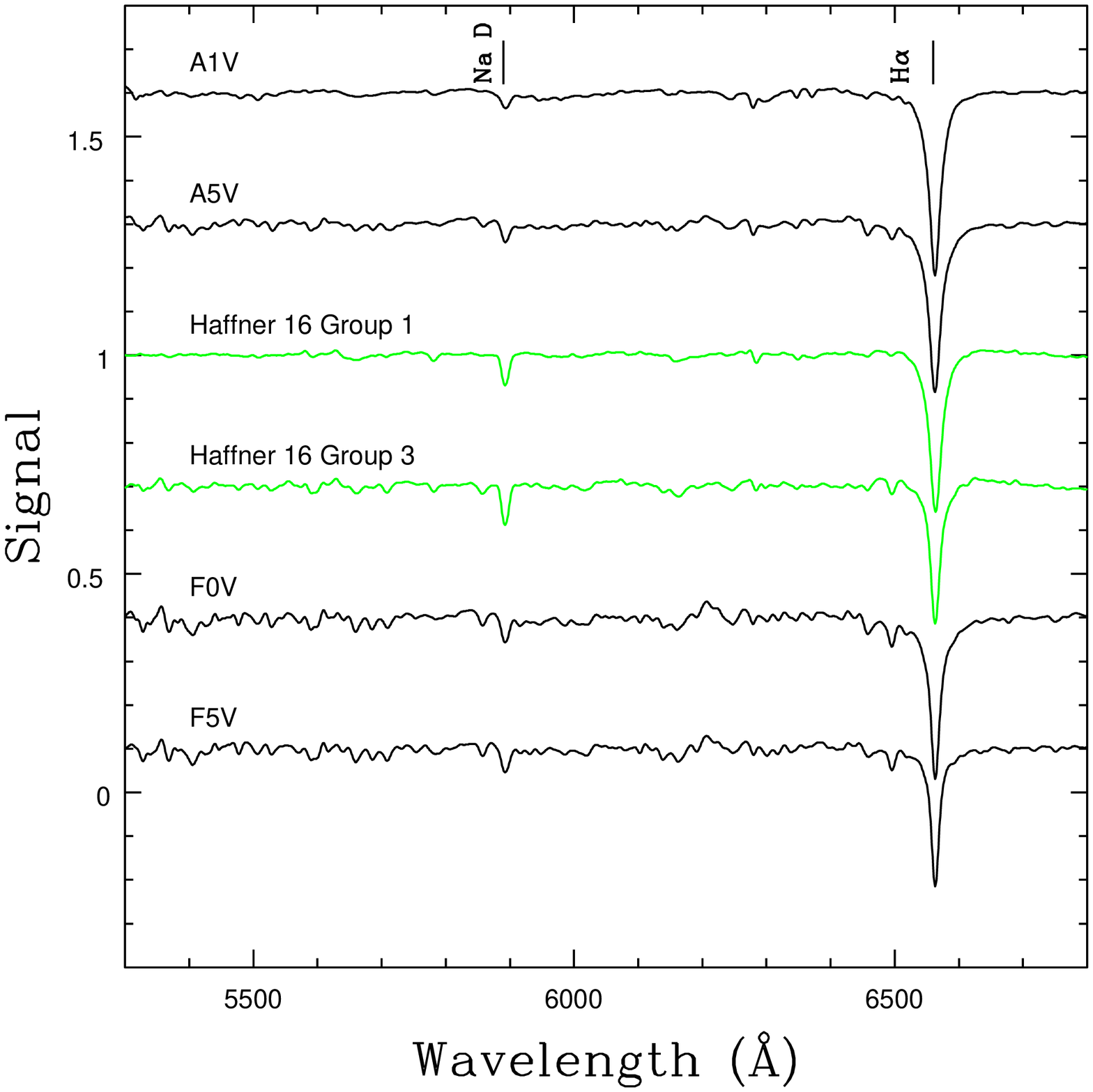}
\caption{Median spectra of stars in groups 1 and 3, which are located near the main 
sequence on the CMD of Haffner 16. Spectra from the Le Borgne et al. (2003) 
library are also shown. All spectra have been smoothed to a resolution of 640 
and normalized to the continuum. The modest depths of metallic absorption features 
in the group 1 and 3 spectra are consistent with early to mid-A spectral-types.}
\end{figure*}

\begin{figure*}
\figurenum{15}
\epsscale{1.0}
\plotone{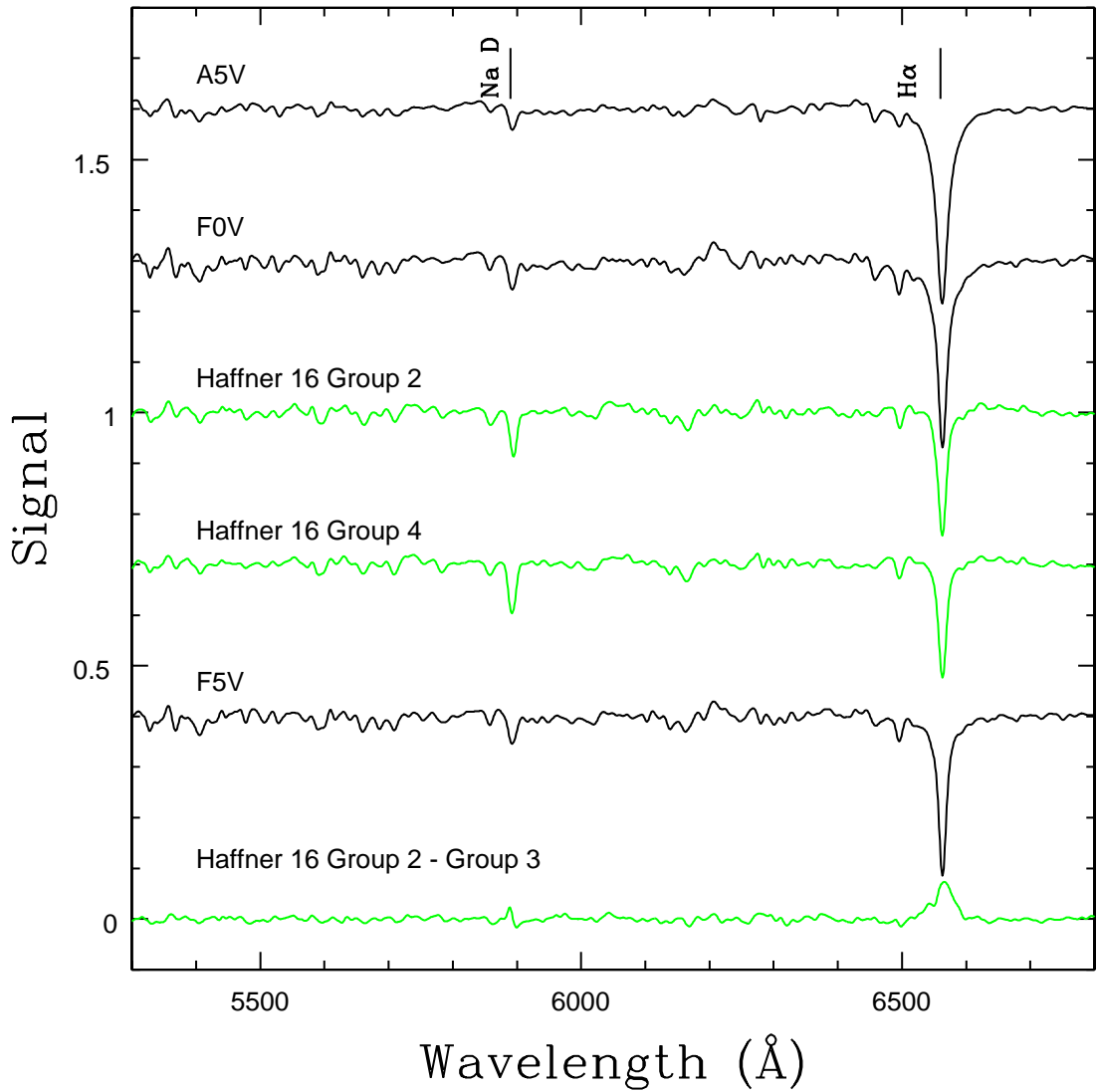}
\caption{Same as Figure 14, but showing the composite spectra of stars in groups 2 and 
4. With the exception of minor differences associated with Na D and H$\alpha$, the 
group 2 and 4 spectra are similar, even though the groups sample stars in different 
magnitude ranges. The metallic lines in these spectra have depths that are consistent 
with solar neighborhood stars that have early to mid-F spectral types. H$\alpha$ has 
a depth that is weaker than expected based on the depths of metallic lines, 
and this is attributed to a sub-solar metallicity.
The difference between the group 2 and 3 spectra is 
also shown. There is a tendency for group 2 to have slightly deeper metallic lines.}
\end{figure*}

\begin{figure*}
\figurenum{16}
\epsscale{1.0}
\plotone{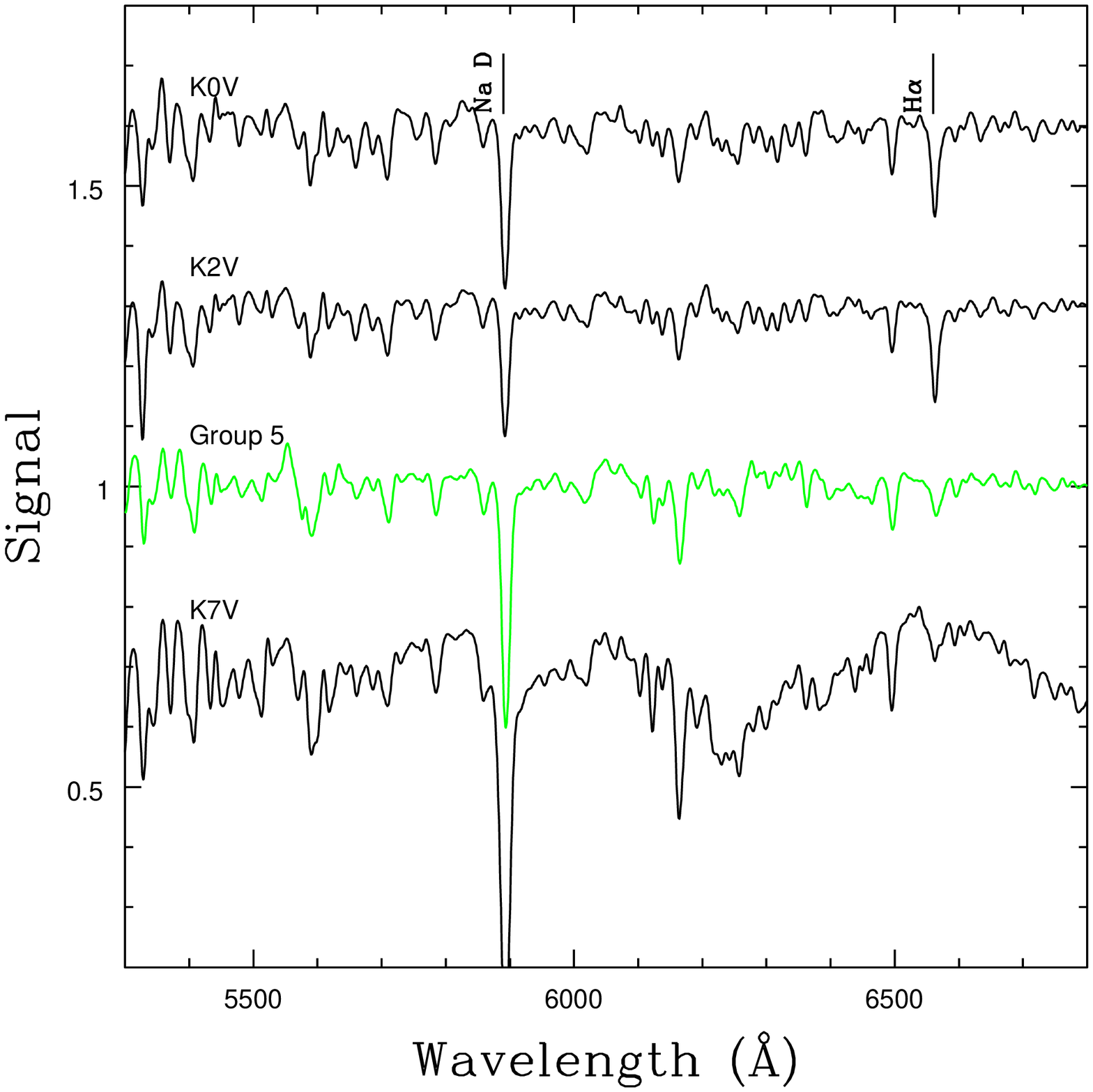}
\caption{Same as Figure 14, but showing stars in group 5. 
Absorption features in the composite spectrum have depths that are 
similar to those in the K0V and K2V spectra.}
\end{figure*}

	With the exception of Na D, which has a contribution from interstellar 
absorption, the depths of features between 5300 and 6500\AA\ in the group 1 
spectrum in Figure 14 are similar to those in the A1V spectrum. 
The overall appearance of the group 3 spectrum between 5300 and 6500\AA\ 
is consistent with a mid A spectral-type. To quantify these comparisons, 
various reference stars were subtracted from the 
group 1 and 3 spectra, and the residuals examined. The 
5400 -- 5800 \AA\ and 6000 -- 6400\AA\ wavelength intervals contain numerous 
metallic features, and the residuals at these wavelengths suggest that 
the characteristic spectral-type of the group 1 spectrum is early to mid-A. 
The characteristic spectral-type for group 3 is A5V. 
However, the depth of H$\alpha$ in the group 3 spectrum 
is more appropriate for an F spectral-type.

	Turning to the spectra of groups 2 and 4 in Figure 15, the depths of 
most metallic lines in the 5300 - 6500\AA\ interval are consistent with early to 
mid-F spectral-types. That the group 2 and 4 spectra are similar indicates that the 
stars in these groups tend to have comparable spectroscopic properties. 
Subtracting the reference star spectra from the group 2 and 4 
spectra and examining the differences in the 5400 -- 5800\AA\ and 
6000 -- 6400\AA\ intervals suggests that an F5V spectral-type characterizes 
both groups. 

	The difference between the mean group 2 and 3 spectra 
is also shown in Figure 15. With the exception 
of Na D and H$\alpha$, the residuals scatter about zero. 
There is a tendency for features to be stronger in the group 2 spectrum when compared 
with the group 3 spectrum, and this is consistent with the difference in spectral 
types inferred from the reference stars.

	The comparisons in Figure 15 indicate that H$\alpha$ is significantly 
shallower in the group 2 and 4 spectra than is seen among solar neighborhood main 
sequence stars with mid-F spectral types. What causes H$\alpha$ to be weaker 
than expected in the group 2 and 4 spectra? This is likely not a surface 
gravity effect, as the majority of stars along this line of sight -- either in 
the field or in Haffner 16 -- are expected to be evolving on the main sequence.
The similarity of the spectra of groups 2 and 4 also indicates that the 
relative strengths of metallic lines and H$\alpha$ are not a 
fluke arising from a peculiar mix of stars. Indeed, 
the influence of stars that may have peculiar properties are suppressed 
by taking the median of spectra in each group.
The relative weakness of H$\alpha$ with respect to the depths of metallic lines 
when compared with solar neighborhood stars is also likely not due 
to velocity smearing. Not only do the stars with spectra in this 
sample have similar velocities, but velocity smearing will affect 
all absorption features, not just H$\alpha$. Finally, H$\alpha$ emission can affect the 
depth of H$\alpha$ absorption, while not affecting the depths of metallic lines. 
However, such emission is not a plausible explanation for the relative weakness of 
H$\alpha$ seen here. This is because groups that sample cluster and 
field members show similar deficiencies in the depth of H$\alpha$. Remarkably similar 
levels of H$\alpha$ emission in the spectra of stars in the cluster and field 
would then have to be present for this to explain the depths of H$\alpha$.

	The shallow depth of H$\alpha$ in the spectra of groups 2 and 4 
is likely a consequence of a property that is common to the stars along the 
Haffner 16 line of sight. That Haffner 16 is located outside of the 
solar circle leads us to suspect that this property is metallicity, in the sense 
that the majority of stars in our spectroscopic sample have a metallicity 
that is lower than that of stars that make up the reference spectra. To understand 
how the depth of H$\alpha$ can be affected by metallicity, consider two hypothetical 
stellar systems, one containing metal-poor stars and the other containing metal-rich 
stars. At a fixed effective temperature the depths of metallic features scale with 
metallicity -- the depths of metallic features in 
the spectra of stars in the metal-poor system will be shallower than those 
in the spectra of stars in the metal-rich system 
that have the same effective temperature. However, the strengths of 
metallic features at visible wavelengths increase as effective temperature decreases. 
Spectra of stars in the metal-poor system will have metallic lines with 
depths that are similar to those seen among stars in the metal-rich system that 
have higher effective temperatures. The Balmer lines are 
less susceptible to metallicity variations as the percentage 
difference in Hydrogen content between the two hypothetical systems is much smaller 
than that between metals, and so the depths of the Balmer lines 
are mainly set by effective temperature -- the depths of Balmer absorption in stars 
in the metal-rich and metal-poor systems that have similar metal line strengths 
will then differ. That H$\alpha$ in the group 3 spectrum is 
shallower than expected based on the strengths of 
metallic features when compared with the solar neighborhood thus suggests that 
Haffner 16 -- like the field stars in groups 2 and 4 -- has a sub-solar 
metallicity.

	The composite spectrum of stars in group 5 is compared with 
those of late-type solar neighborhood stars in Figure 16. The majority of 
metallic features in the Group 5 spectrum have depths that match those in the K0V and 
K2V spectra. The absence of TiO absorption, which is seen in the 
K7V spectrum between 6100 and 6500\AA, is consistent with a spectral-type 
not later than mid-K. Various reference spectra 
were subtracted from the group 5 spectrum, and the residuals 
in the 5400 -- 5800\AA\ and 6000 --- 6400\AA\ intervals are consistent with a K2V 
characteristic spectral-type for the group 5 spectrum. 
As was seen amongst the brighter stars, H$\alpha$ absorption 
in the group 5 spectrum is weaker than expected based on the 
depths of metallic features, again hinting at a sub-solar metallicity.

	A sub-sample of stars have spectra that extend to 4800\AA, and these are 
objects that are located to the left of the cluster center in Figure 1. 
The wavelength range between 4800 and 5300\AA\ is of interest as it contains 
prominent features that probe metallicity and surface gravity, including H$\beta$, 
Mgb, and Fe 5270\AA. Median spectra of the stars that have 
H$\alpha$ in absorption and that have spectra extending down to 4800\AA\ 
are shown in Figures 17, 18, and 19. The spectra in these figures have been smoothed 
to a resolution of 640 and have been normalized to the continuum.

\begin{figure*}
\figurenum{17}
\epsscale{1.0}
\plotone{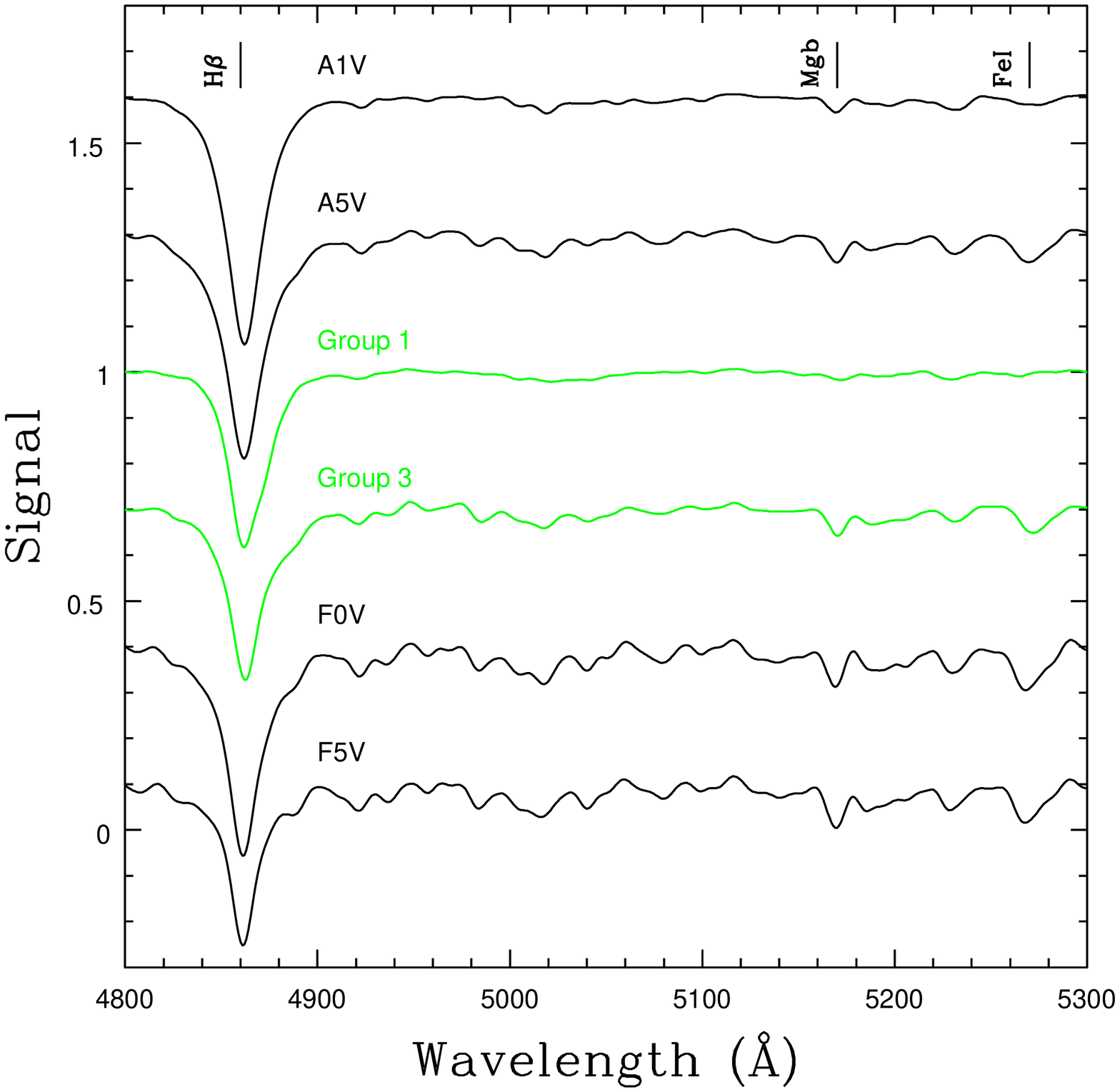}
\caption{Median spectra of stars in groups 1 and 3 with wavelength coverage 
that extends to 4800\AA. Spectra from the Le Borgne et al. (2003) library are also 
shown. The spectra have been smoothed to a resolution of 640 and 
normalized to the continuum. The depths of metallic absorption 
features in the Haffner 16 spectra are consistent with those in early to mid-A 
spectral types. However, H$\beta$ is shallower than expected for an 
A-type main sequence star.}
\end{figure*}

\begin{figure*}
\figurenum{18}
\epsscale{1.0}
\plotone{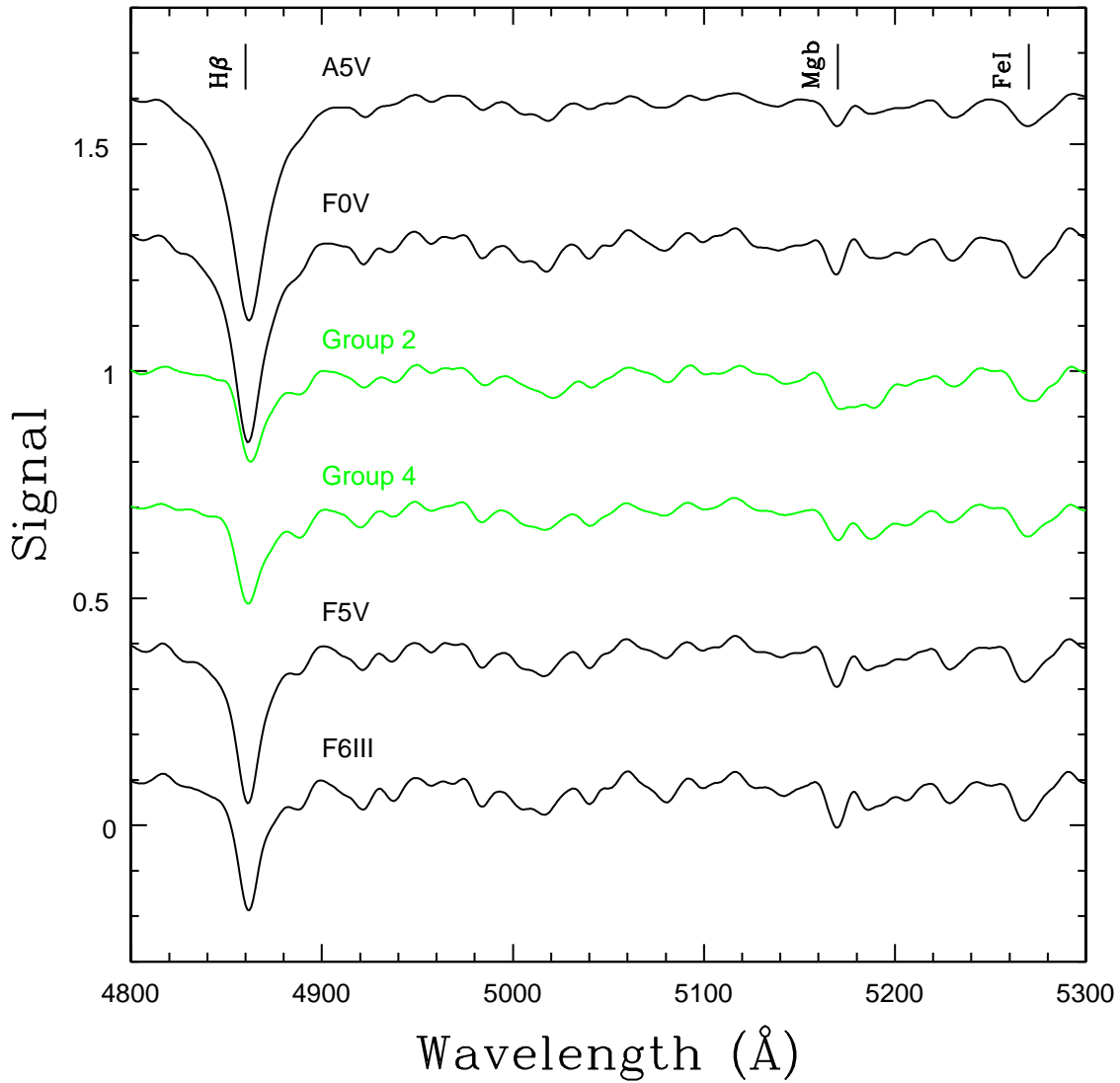}
\caption{Same as Figure 17, but showing the median spectra of the 
stars in groups 2 and 4 with wavelength coverage extending to 4800\AA. 
As at longer wavelengths, the median spectra of these groups are very similar, 
with metallic line strengths that match those in solar neighborhood 
stars that have a late-A or early F spectral type. However, the depth of H$\beta$ is 
consistent with later spectral-types.}
\end{figure*}

\begin{figure*}
\figurenum{19}
\epsscale{1.0}
\plotone{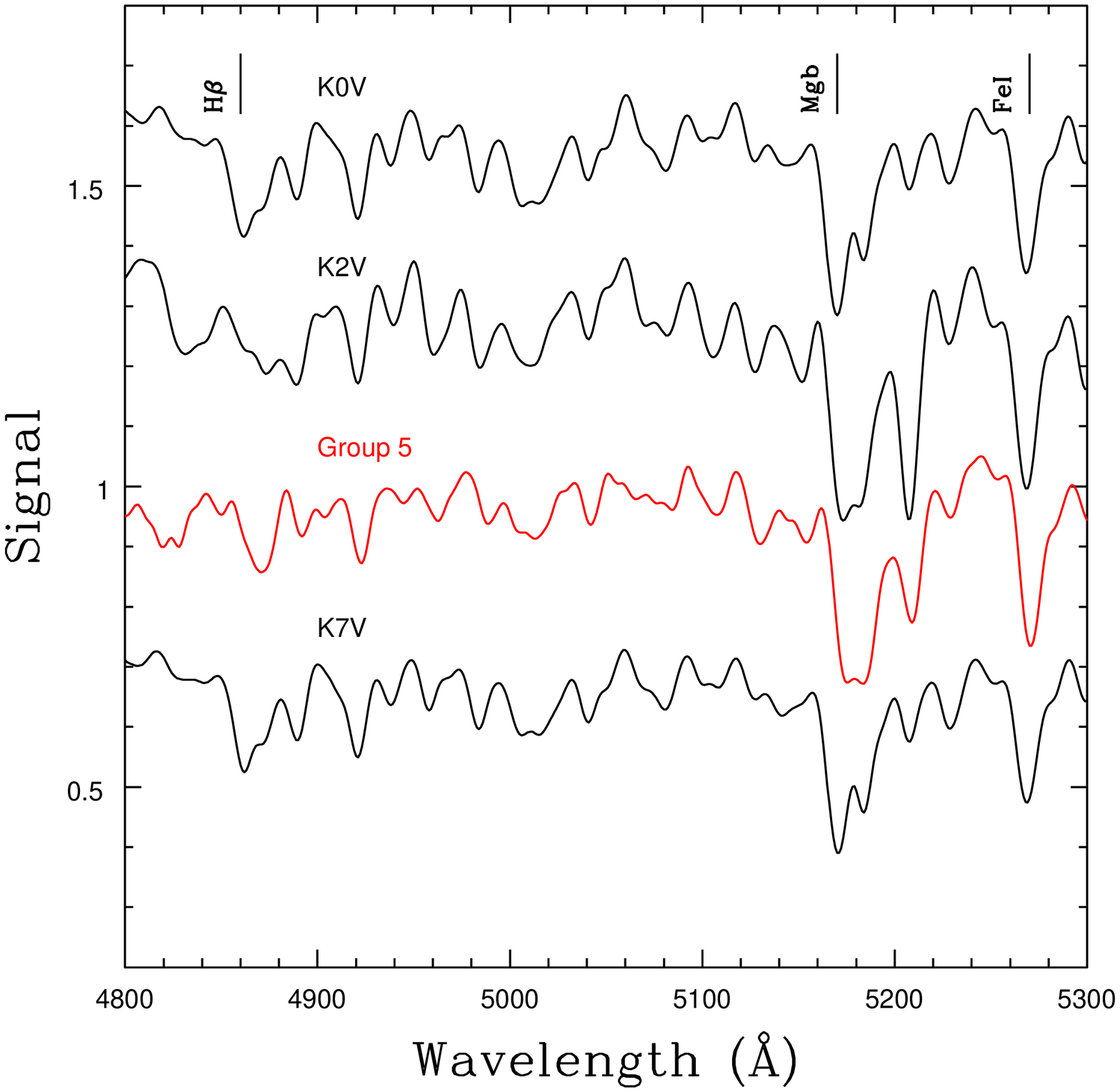}
\caption{Same as Figure 17, but showing stars in group 5. 
The group 5 spectrum is consistent with a K spectral type.}
\end{figure*}

	The spectra in Figures 17, 18, and 19 were 
constructed by combining only a quarter to a third of the total number of objects 
that were used to construct the spectra in Figures 14, 15, and 16. 
The comparisons in these figures thus not only provide information about 
wavelength-related effects but also the sensitivity of the results gleaned 
at longer wavelengths to sample size. It is thus worth noting that 
comparisons with the depths of metallic features in the Le Borgne et al. (2003) 
spectra at blue wavelengths yield characteristic spectral-types that are similar 
to those found at longer wavelengths. In all five groups H$\beta$ is also shallower 
than expected based on the strengths of the metallic features between 
4900 and 5300\AA\ when compared with the Le Borgne et al. spectra. 
The comparisons involving spectra between 4800 and 5300\AA\ 
thus reinforce the results found from spectra at longer wavelengths.

\subsection{H$\alpha$ Emission Stars}

	Composite spectra of stars in groups 6 and 7 are shown in Figure 20. 
There are similarities between the spectra in the wavelength interval 5300 -- 5800\AA. 
However, the group 6 and 7 spectra differ at wavelengths $> 5800\AA$. 
There are differences in the depths of Na D, as well as 
the Ca I feature at 6170\AA. Comparisons with spectra of K2V and K7V stars from Le 
Borgne et al. (2003), also shown in Figure 20, suggest that the 
depths of metallic lines in these spectra are consistent with an early to 
mid-K spectral type. The characteristic spectral types of the group 6 and 7 
spectra were estimated by subtracting various reference star 
spectra from the group 6 and 7 spectra, and then examining 
the residuals in the 5400 -- 5800\AA\ and 6000 -- 6400\AA\ intervals. 
The K2V reference star gives the best match to the group 6 and 7 spectra.

\begin{figure*}
\figurenum{20}
\epsscale{1.0}
\plotone{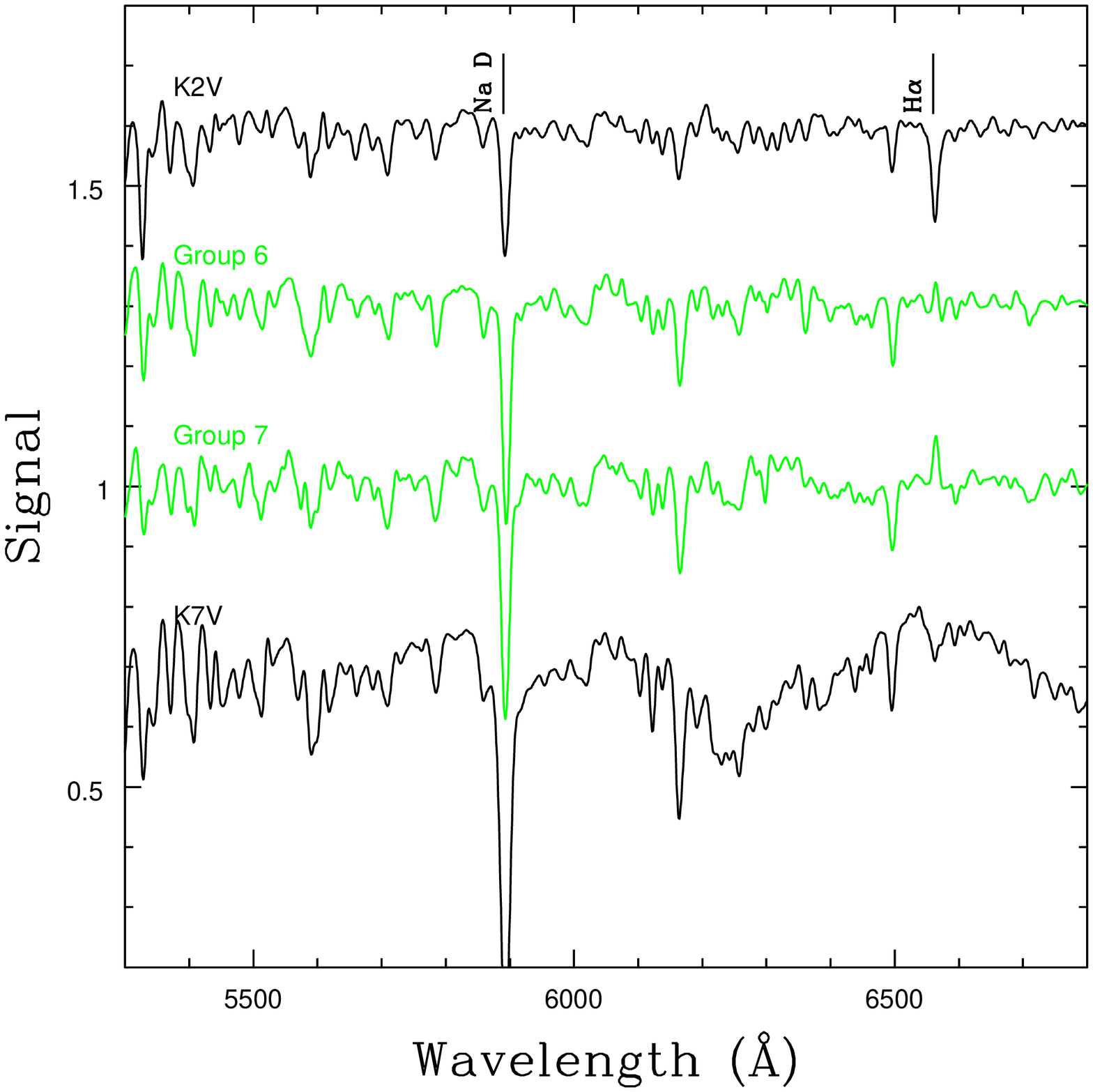}
\caption{Median spectra of stars in groups 6 and 7 are compared with 
spectra of stars from the Le Borgne et al. (2003) library. Na D and Ca I 6170\AA\ in 
the group 7 (stars with pronounced H$\alpha$ emission) 
spectrum are deeper than in the group 6 (stars with weak H$\alpha$ 
emission) spectrum. There are also subtle differences between the group 6 and 7 spectra 
at wavelengths $> 6200$\AA. The behaviour of H$\alpha$ aside, the depths of metallic 
features in both groups are consistent with an early to mid-K spectral-type.}
\end{figure*}

	While the spectral characteristics of groups 6 and 7 are broadly 
similar to those of group 5, there are differences, and these are 
examined in Figure 21, where the results of subtracting the group 5 spectrum from the 
group 6 and 7 spectra are shown. With the exception of the depth of Na D, 
the group 6 spectrum is an overall better match to the group 5 spectrum than the 
group 7 spectrum. However, the difference in Na D depth is such that Na D is 
stronger in group 5 than in group 6, and the Na D line in the group 7 spectrum 
more closely matches to that in the group 5 spectrum than in group 6. Thus, 
the strength of H$\alpha$ emission is not related to Na D strength.

\begin{figure*}
\figurenum{21}
\epsscale{1.0}
\plotone{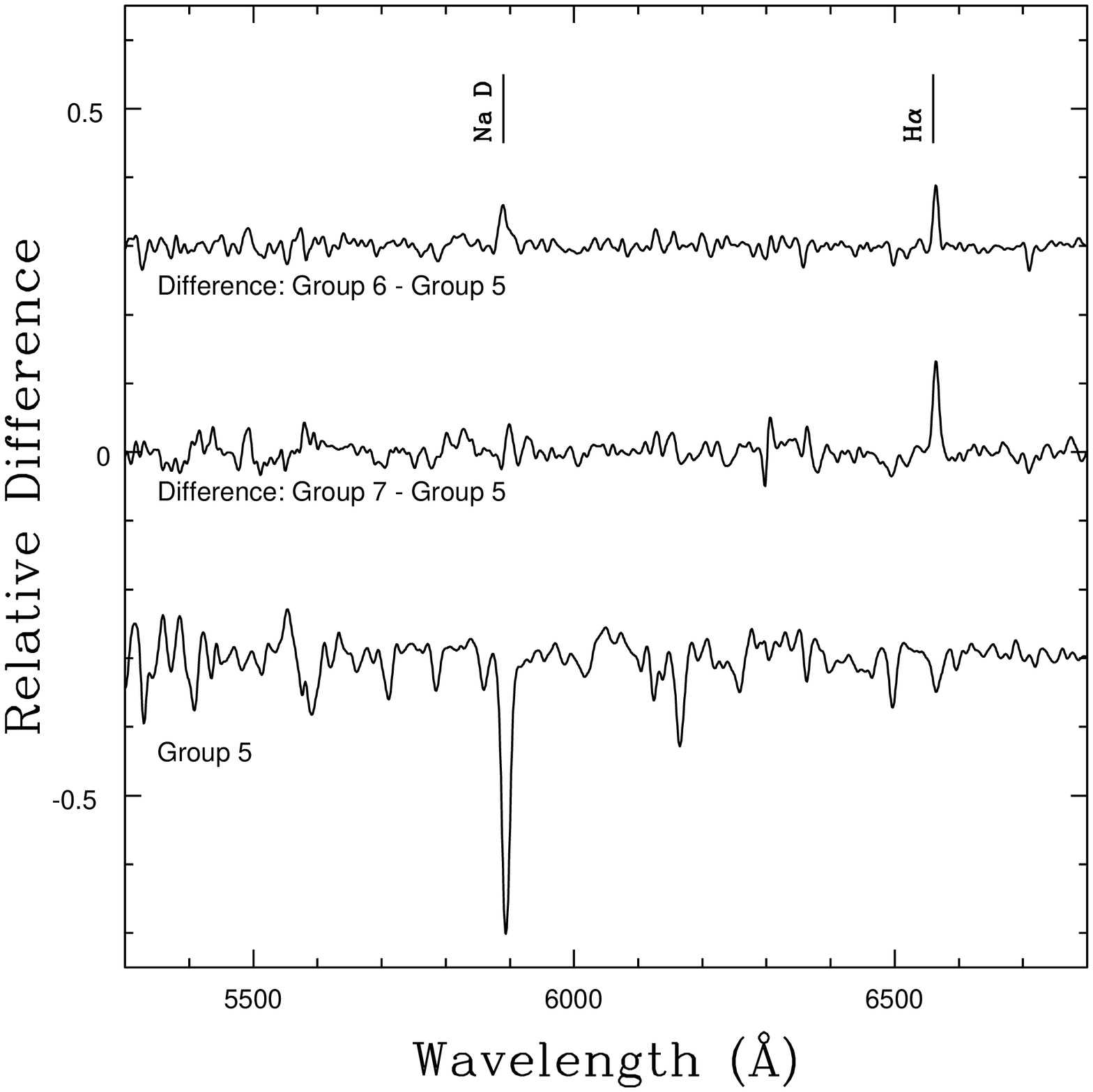}
\caption{Differences between the spectra of the two groups with H$\alpha$ 
in emission and the group 5 spectrum. The group 5 spectrum from 
Figure 16 is shown to assist with line identifications. With the exception of 
Na D, the residuals indicate that the group 5 spectrum is better matched by the group 
6 spectrum than the group 7 spectrum. Any differences in line strength due to 
circumstellar absorption must be modest given the similar mean $g'-i'$ colors 
of stars in the groups (see text).}
\end{figure*}

	The mean colors of stars in groups 5, 6, and 7 are not significantly different. 
Stars in group 7 have a mean $g'-i' = 1.80 \pm 0.06$, 
whereas the mean for stars in group 6 is $1.79 \pm 0.05$. The uncertainties 
are the standard error about the mean. The similarity in color makes 
it unlikely that the differences in the depth of Na D between group 6 and 7 
is due to circumstellar absorption by a dust-rich component. The equivalent width of 
H$\alpha$ emission also appears not to be related to $g'-i'$ color.

	A moderate sub-sample of the stars in group 6 have 
wavelength coverage that extends to 4800\AA, and the median spectrum of 
sources in group 6 with blue wavelength coverage is shown in Figure 22. 
Also shown is the spectrum of a K2V star and the median 
spectrum of sources in group 5 from Figure 19. Unfortunately, there are not 
enough stars in Group 7 with wavelength coverage between 4800 and 5300\AA\ 
to construct a spectrum with even a moderately high S/N ratio. 

\begin{figure*}
\figurenum{22}
\epsscale{1.0}
\plotone{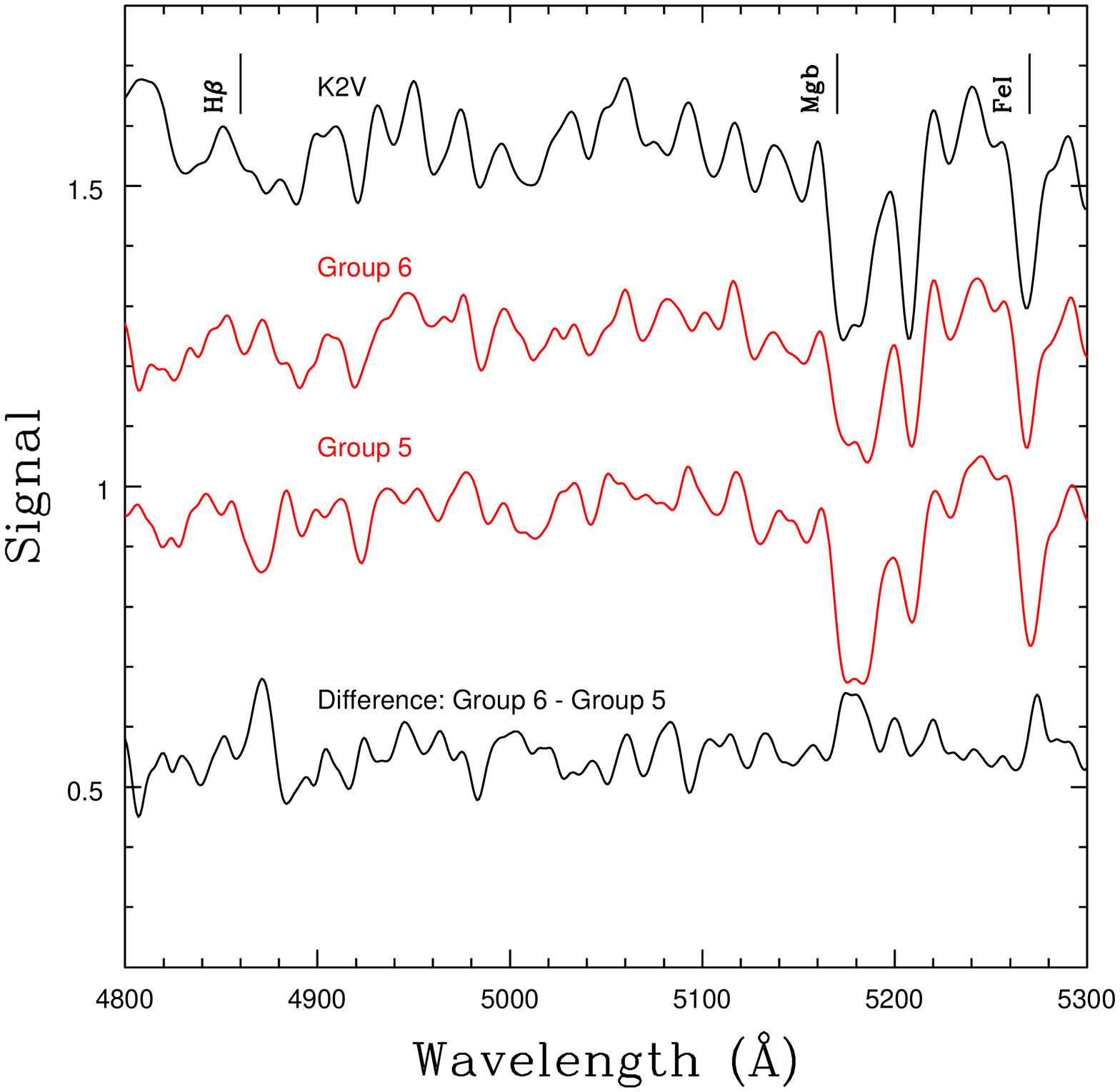}
\caption{Spectra of groups 5 and 6 at blue wavelengths. 
The spectrum of a K2V star from the Le Borgne et al. (2003) library is also 
shown. The difference between the group 5 and 6 spectra shows larger residuals 
than when the spectra at longer wavelengths are compared. This is 
due in part to smaller sample sizes and the comparatively low system throughput 
at wavelengths $< 5300$\AA. Still, there are indications that absorption features 
in group 6 at these wavelengths are weaker than those in group 5.}
\end{figure*}

	The difference between the group 6 and group 5 spectra is shown 
in Figure 22. The scatter in the residuals is larger than 
in Figure 21. Only a small number of sources have been 
combined to construct the spectra in Figure 22, and there is lower optical 
throughput at blue wavelengths than at longer wavelengths. 
The S/N ratio of the blue spectra in Figure 22 is thus lower than at longer 
wavelengths. There is a possible difference in H$\beta$ strength between 
the two groups, although the statistical significance is marginal. 
This might indicate H$\beta$ emission in the group 6 spectrum, 
or could simply indicate a difference in the depth of H$\beta$ absorption. 
Fe I 5270 and Mgb both appear to be weaker 
in the group 6 spectrum than in the group 5 spectrum. 
The group 6 spectum in Figure 22 is consistent with a K 
spectral type, which is in line with the comparisons made at longer wavelengths.

\subsection{Spectra of Four Serendipitous Targets}

	The spectra of the four sources that were observed (1) because they could 
be placed in the same slit as a CMD-selected target, and (2) have locations on 
the CMD that are distinct from the 7 groups defined above are shown in Figure 23. 
The locations of 4-21b and 6-02a in the GMOS science field are such that spectra 
at wavelengths $< 5800\AA$ were not recorded, and so the spectra in Figure 23 are 
restricted to the wavelength interval 5800 -- 6800\AA. The feature near 6300\AA\ in 
each spectrum is an artifact of [OI] airglow subtraction -- these artifacts are not 
seen in the group 1 -- 7 spectra as they are suppressed when individual spectra are 
combined together. The 5-09b spectrum is nulled between 6300 and 6400 \AA\ because 
this interval samples the gap between CCDs, and the signal was excessively noisy 
there. Comments on the spectra are as follows:

\begin{figure*}
\figurenum{23}
\epsscale{1.0}
\plotone{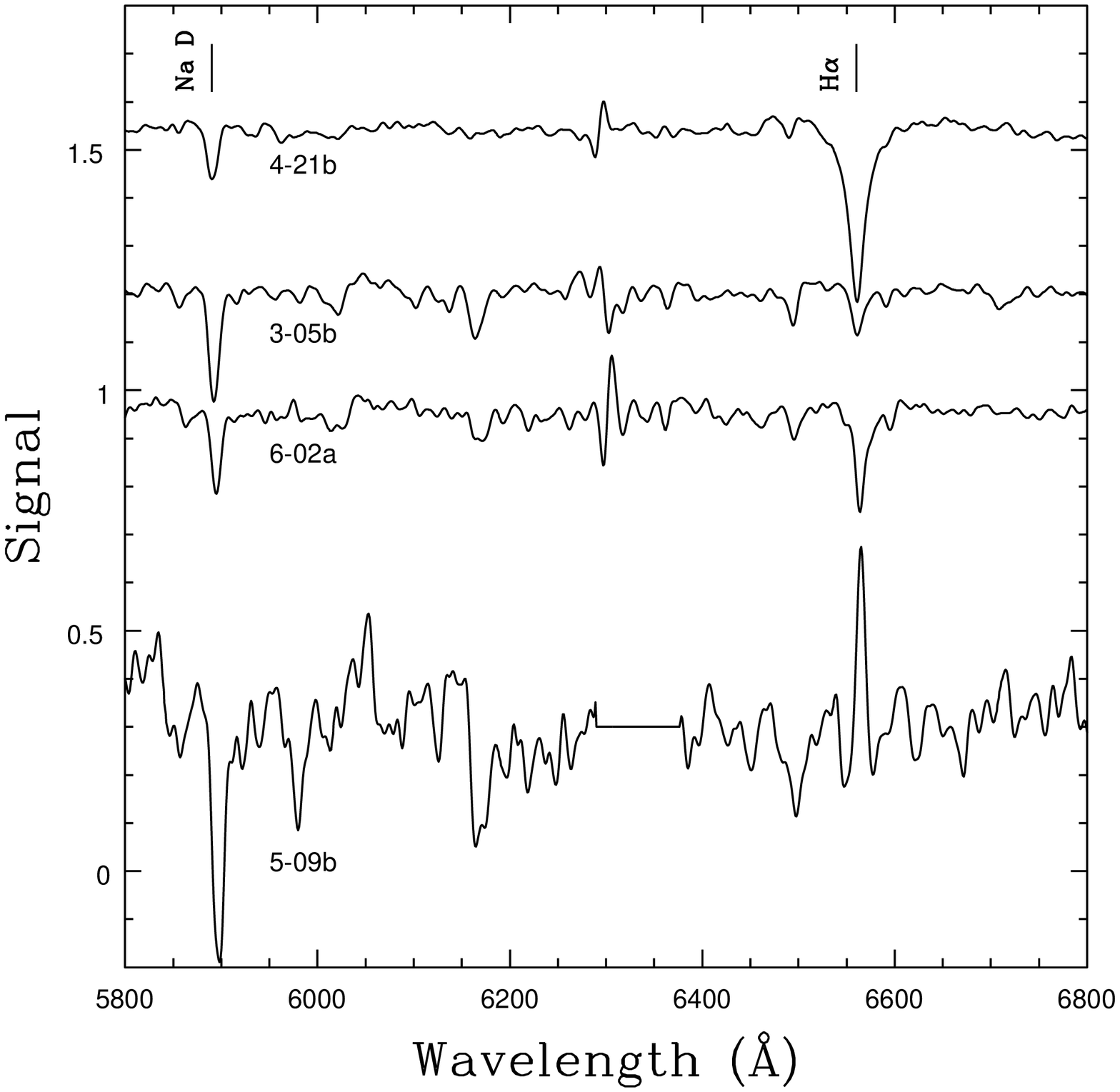}
\caption{Spectra of four stars in the serendipitous sample. 
The locations of these stars on the CMD are indicated in 
Figure 10. The spectra have been smoothed to a resolution 
of 640, and normalized to the continuum. The noise feature near 6300\AA\ in each 
spectrum is an artifact of [OI] airglow emission. Star 5-09b is the faintest object 
in the spectroscopic sample, and the region that is nulled between 6300 and 6400\AA\ in 
its spectrum is the gap between CCDs, where the S/N ratio is poor.
It is argued in the text that 4-21b and 3-05b are probably not 
cluster members, while the membership of 6-02a is uncertain. 
However, the photometric and spectroscopic properties of 
5-09b suggests that it may belong to Haffner 16.}
\end{figure*}

\noindent{\bf 4-21b:} The location of this star on the CMD places it $\sim 0.4$ 
magnitudes in $g'-i'$ blueward of the cluster sequence, close 
to the near-vertical blue field star sequence. These photometric 
properties suggest that 4-21b almost certainly does not belong to Haffner 16. 
Na D in the 4-21b spectrum is deeper than in groups 1 and 3, while
H$\alpha$ is comparable in depth to that in the group 1 spectrum.
4-21b is probably a background F star.

\noindent{\bf 3-05b:} The $g'-i'$ color of star 3-05b places it $\sim 0.5$ magnitude 
redward of the cluster sequence in Figure 10, in a part of the CMD that contains 
few stars. The weak H$\alpha$ absorption suggests a late spectral-type, and this 
is consistent with the deep Na D lines. It is likely that 3-05b is a foreground dwarf.

\noindent{\bf 6-02a:} Star 6-02a falls close to the cluster sequence in Figure 10.
H$\alpha$ absorption is shallower than in the group 3 spectrum, but is comparable 
to that in the group 2 and 4 spectra. The Na D lines are deeper than in the 
spectra of groups 1 -- 4, suggesting a later spectral type than a 
typical star in these groups. With the exception of 
Na D, the metallic lines in the 6-02a and 3-05b spectra have similar depths, and 
this similarity was confirmed by examining the difference between the 
two spectra. While the $g'-i'$ color of 6-02a is close to that expected for a 
member of Haffner 16 at this brightness, this star also falls squarely on the 
foreground dwarf sequence in the CMD. Thus, the relationship between 6-02a and 
Haffner 16 is uncertain.

\noindent{\bf 5-09b:} Star 5-09b is the faintest source for which a spectrum was 
recorded, and there is considerable noise in the spectrum, 
even after smoothing. 5-09b falls just redward 
of the cluster sequence in Figure 10, with an offset in $g'-i'$ 
that does not exceed those of stars in groups 5, 6, and 7. The Na D lines are 
deeper than in the composite spectra of groups 6 and 7, suggesting 
a later spectral-type than the majority of stars in those groups. 
The equivalent width of H$\alpha$ emission in the 5-09b spectrum is by far the largest 
of any star in the spectroscopic sample. These characteristics suggest that 5-09b is a 
young K dwarf, and is a possible PMS star that belongs to Haffner 16.

\section{DISCUSSION \& SUMMARY}

	Images and spectra recorded with GMOS on Gemini South have been used to 
examine the spectrophotometric properties and spatial distributions of stars in 
the young open cluster Haffner 16. A tight, well-populated 
sequence is found in the CMD of objects within 90 arcsec of the cluster center, 
and this enables the identification of candidate cluster members that span a 
range of masses. The photometric properties of sub-solar mass 
stars in Haffner 16 is of particular interest as these objects can provide 
constraints on the ages of young clusters. Studies of the sub-solar mass 
regime in young clusters might also provide insights into the photometric 
properties of PMS stars and thereby set constraints for the models that seek to 
replicate their properties. 

	The age estimated for Haffner 16 from the $(i', g'-i')$ CMD and $i'$ LF is 
older than that found by Davidge et al. (2013) from a deep $(K, J-K)$ CMD and $K$ LF. 
Both sets of data track the cluster PMS down to masses of a few tenths solar. 
The difference in age estimates is due to the distance moduli adopted 
-- the Davidge et al. (2013) age of $< 10$ Myr assumes a 
distance modulus of 13.5, while the CMD and LF constructed from the present data 
support an age of 20 Myr with a distance modulus $\sim 12.3$. The age found 
here is preferable as the visible/red CMD obtained from the GMOS images provides 
firmer constraints on the distance of Haffner 16 than the NIR CMD. 
Indeed, structures such as the bend in the CMD of MS stars near $i' = 15.5$ 
that serve as anchor points for isochrones are more pronounced at visible 
wavelengths than in the NIR.

	The age found here is older than that estimated by McSwain 
\& Gies (2005), even though that study assumed a distance modulus of 12.5. 
The disagreement in age estimates in likely tied to differences in the area sampled 
on the sky. The present study considers only objects that are 
within 1.5 arcmin from the cluster center, which 
the cluster light profile indicates is an area that is dominated 
by cluster stars. In contrast, McSwain \& Gies consider 
objects over a 13.5 $\times$ 13.5 arcmin area, with the result that 
there is substantial contamination from field stars. While a modest population of stars 
related to Haffner 16 {\it may} be present in this area (Figure 2), 
there is no guarantee that these are cluster members or that they formed at the 
same time as the cluster.

	Differences in the age estimates notwithstanding, the line of sight reddening 
found here agrees with that measured by Davidge et al. (2013). 
Reddening estimates that are based on the photometric properties of early-type MS stars 
in Haffner 16 are not sensitive to the adopted distance as these objects 
define a more-or-less vertical trajectory on the visible and NIR CMDs. The agreement 
between extinction estimates made over a large wavelength range validates the 
reddening law used in this study, which is the R$_V = 3.1$ relation from Cardelli et 
al. (1989), as summarized in Table 6 of Schlegel et al. (1998). 

	Haffner 16 has a well-populated PMS, and so is an interesting laboratory 
for testing models of PMS evolution. Past studies of young clusters 
have shown that the agreement between the observed photometric 
properties of PMS stars and those predicted by models depends on wavelength. 
Bell et al. (2012) investigated the agreement between models and 
observations of PMS stars in the Pleiades, which is many tens of 
Myr older than Haffner 16. While good agreement was found 
between models and observations in $K$, this was not the case at 
wavelengths shortward of $1\mu$m. Similar behaviour was found by Davidge (2014; 2015)
when comparing isochrones with the $(i', g'-i')$ and $(K, J-K)$ CMDs of the 20 Myr 
cluster NGC 2401. 

	As in the Pleiades and NGC 2401, the level of agreement between the models and 
observations for Haffner 16 depends on wavelength. Davidge et al. (2013) found 
that solar metallicity isochrones matched the location of PMS stars in Haffner 16 on 
the $(K, J-K)$ CMD over the full range of magnitudes sampled. The PMS sequence 
in the NIR CMD is almost vertical, and so is not 
sensitive to the adopted distance modulus. However, 
in Section 5 it is shown that the same isochrones do not track the PMS sequence 
on the $(i', g'-i')$ CMD, falling progressively blueward of cluster 
stars when $i' < 18$. 

	What is the cause of the offset between the models 
and the Haffner 16 sequence on the $(i', g'-i')$ CMD? 
The presence of line emission among PMS stars in Haffner 16 may provide 
clues to answering this question. Davidge et al. (2013) discuss narrow-band images 
that reveal possible Br$\gamma$ emission among stars that have $K > 15$ in Haffner 16. 
If Br$\gamma$ emission is present then prominent H$\alpha$ emission should also be seen 
among faint stars in Haffner 16, and the GMOS spectra confirm that this is the case. 
That the majority of H$\alpha$ emission sources tend to be found within or near 
the Shoulder region (Figure 12) is consistent with them being cluster members. 

	Br$\gamma$ and H$\alpha$ emission are signatures of chromospheric activity. 
Such activity among PMS stars in Haffner 16 might occur if these objects 
are accreting material. Obscuration by circumstellar material might then
explain the red colors of PMS stars, although this would require a remarkable 
alignment of accretion disks along the line of sight, coupled with a level of 
obscuration that increases towards low masses. However, accretion is an unlikely 
explanation for line emission in Haffner 16, as the time scale for the decay of 
stellar accretion disks is a few Myr (e.g. Haisch et al. 2001; Fedele et al. 2010, but 
see also De Marchi et al. 2011; 2013), whereas Haffner 16 has an age of $\sim 20$ Myr.

	Strong emission lines and associated continuum emission will affect 
broad-band colors. However, line and continuum emission do not
affect the $g'-i'$ colors of PMS stars in Haffner 16 by a significant amount. 
Evidence for this comes from the $g'-i'$ colors of the stars in 
groups 5, 6, and 7. The mean colors of stars in these 
groups are similar, even though they have very different 
H$\alpha$ characteristics (Section 7.3). The similar metallic line strengths seen 
in the composite spectra of these groups also suggests that line veiling -- which will 
occur if there is continuum emission -- is not significant in the stars that have 
H$\alpha$ in emission.

	Star spots with temperatures that are cooler than the surrounding photosphere 
are a possible explanation of the offets between the models and 
observations. Spot activity can be driven by rapid 
rotation in young stars, and has been proposed to explain 
the SED of low mass stars in the Pleiades (e.g. Stauffer et al. 2003). 
Somers \& Pinsonneault (2015) examine the influence that cool star spots 
may have on the photometric properties of PMS stars. They consider spots that 
have properties similar to those seen in actual stars, and investigate the 
affect that spots have on photometric properties with respect to stellar 
mass (and hence effective temperature). Their models show that 
light from stars at visible wavelengths is susceptible to spot activity as this 
is where the SEDs of spots peak; in contrast, the NIR part of the 
spectrum is little changed. Star spots can thus explain qualitatively the 
offsets between isochrones and the Haffner 16 PMS at visible wavelengths, and 
the agreement between observations and theory in the NIR.

	The majority of candidate PMS stars with spectra in this study have $i' \sim 
18$, and there is only a small offset between the isochrones and observed PMS sequence 
at this magnitude (e.g. Figure 5). While H$\alpha$ emission is seen at this 
brightness, it has a modest strength. If spots affect 
the broad-band colors of PMS stars in Haffner 16 then 
spectra of fainter (and thus lower mass) PMS stars in this cluster should show 
more pronounced H$\alpha$ emission, that correlates with the difference in 
$g'-i'$ between the isochrones and observations. 
This expectation is consistent with the level of H$\alpha$ emission seen in 
the spectrum of 5-09b, which is the faintest source with a spectrum in this study, 
and is also the object in the current sample with the 
largest equivalent width of H$\alpha$ emission. Spectra or narrow-band H$\alpha$ 
images of sources with $i' > 19$ in Haffner 16 will reveal if the spectroscopic 
properties of 5-09b are typical of faint stars in the cluster. If spots do affect the 
photometric properties of PMS stars in Haffner 16 
then high dispersion spectra should also reveal these stars to be rapid rotators. 
The moderately tight nature of the PMS in the CMD would require that the majority 
of cluster stars be spotted. By extension, spectroscopic and/or narrow-band 
observations of PMS stars in the 20 Myr old cluster NGC 2401 -- which have 
photometric properties that are similar to their counterparts in Haffner 16 -- 
should also show evidence for prominent H$\alpha$ emission.

\acknowledgements{It is a pleasure to thank the anonymous referee for providing 
timely and comprehensive reports that greatly improved the manuscript.}

\LongTables
\begin{deluxetable*}{rrrrrrr}
\tablecaption{Spectroscopic Targets}
\startdata
\tableline\tableline
ID & RA & Dec & x & y & $i'$ & $g'-i'$ \\
\tableline
1-01*(2-01)	 & 117.543569 & -25.473770 &2038.52 &195.710 &14.591 &-0.042 \\
1-02*(2-02)	 & 117.561257 & -25.483789 &2285.60 &589.080 &14.825 &0.186 \\
1-03	 & 117.566457 & -25.483040 &2267.18 &705.049 &14.596 &0.123 \\
1-04	 & 117.574332 & -25.495831 &2582.41 &879.969 &15.025 &0.313 \\
1-05*(2-05)	 & 117.577887 & -25.443279 &1287.63 &959.210 &14.087 &-0.140 \\
1-06	 & 117.582393 & -25.451210 &1483.15 &1059.30 &14.264 &-0.038 \\
1-07	 & 117.585096 & -25.466511 &1860.14 &1120.01 &14.425 &0.007 \\
1-08	 & 117.588058 & -25.456129 &1604.39 &1185.65 &15.006 &0.050 \\
1-09*(2-09)	 & 117.592056 & -25.478230 &2149.06 &1274.56 &14.662 &-0.124 \\
1-10*(2-10)	 & 117.594252 & -25.450649 &1469.45 &1323.00 &13.650 &-0.137 \\
1-11*(2-11)	 & 117.597878 & -25.459641 &1691.00 &1404.01 &14.571 &-0.092 \\
1-12	 & 117.600510 & -25.478060 &2144.87 &1462.51 &15.024 &-0.140 \\
1-13*(2-13)	 & 117.604551 & -25.452379 &1512.29 &1552.50 &14.831 &0.518 \\
1-14*(2-14)	 & 117.615952 & -25.475849 &2090.76 &1806.00 &14.323 &-0.257 \\
1-15*(2-15)	 & 117.619822 & -25.433611 &1050.01 &1892.39 &14.543 &0.759 \\
1-16*(2-16)	 & 117.630665 & -25.449350 &1437.84 &2134.00 &14.186 &0.617 \\
2-01*(1-01)	 & 117.543569 & -25.473770 &2038.52 &195.710 &14.591 &-0.042 \\
2-02*(1-02)	 & 117.561257 & -25.483789 &2285.60 &589.080 &14.825 &0.186 \\
2-03	 & 117.567780 & -25.452490 &1514.44 &734.239 &15.015 &0.035 \\
2-04	 & 117.572773 & -25.497339 &2619.65 &845.320 &14.906 &0.442 \\
2-05*(1-05)	 & 117.577887 & -25.443279 &1287.63 &959.210 &14.087 &-0.140 \\
2-06	 & 117.583258 & -25.458920 &1672.97 &1078.65 &14.803 &0.210 \\
2-07	 & 117.586255 & -25.457600 &1640.48 &1145.40 &14.810 &-0.120 \\
2-08	 & 117.588887 & -25.439720 &1200.13 &1204.06 &14.639 &0.821 \\
2-09*(1-09)	 & 117.592056 & -25.478230 &2149.06 &1274.56 &14.662 &-0.124 \\
2-10*(1-10)	 & 117.594252 & -25.450649 &1469.45 &1323.00 &13.650 &-0.137 \\
2-11*(1-11)	 & 117.597878 & -25.459641 &1691.00 &1404.01 &14.571 &-0.092 \\
2-12	 & 117.601111 & -25.468460 &1908.41 &1475.93 &14.342 &0.576 \\
2-13*(1-13)	 & 117.604551 & -25.452379 &1512.29 &1552.50 &14.831 &0.518 \\
2-14*(1-14)	 & 117.615952 & -25.475849 &2090.76 &1806.00 &14.323 &-0.257 \\
2-15*(1-15)	 & 117.619822 & -25.433611 &1050.01 &1892.39 &14.543 &0.759 \\
2-16*(1-16)	 & 117.630665 & -25.449350 &1437.84 &2134.00 &14.186 &0.617 \\
3-01*(4-01)	 & 117.544506 & -25.468800 &1915.97 &216.179 &15.472 &0.658 \\
3-02	 & 117.549083 & -25.436319 &1115.77 &317.829 &15.287 &0.807 \\
3-03	 & 117.553968 & -25.488550 &2402.67 &426.859 &15.426 &0.654 \\
3-04*(4-04)	 & 117.556415 & -25.449940 &1451.59 &481.209 &15.420 &0.927 \\
3-05a	 & 117.558746 & -25.461929 &1747.00 &533.239 &15.700 &0.390 \\
3-05b	 & 117.559249 & -25.461905 &1747.60 &550.700&16.847 &1.424 \\
3-06*(4-06)	 & 117.561121 & -25.449490 &1440.56 &585.820 &15.961 &0.702 \\
3-07*(4-07)	 & 117.563975 & -25.453800 &1546.66 &649.270 &15.734 &0.476 \\
3-08	 & 117.567451 & -25.429319 &943.650 &726.770 &15.904 &0.824 \\
3-09*(4-09)	 & 117.570255 & -25.452900 &1524.51 &789.250 &16.015 &0.542 \\
3-10	 & 117.575862 & -25.471861 &1991.70 &914.030 &15.194 &0.577 \\
3-11	 & 117.580469 & -25.452280 &1509.52 &1016.55 &15.571 &0.596 \\
3-12	 & 117.587435 & -25.445601 &1344.93 &1171.42 &15.159 &0.619 \\
3-13	 & 117.591949 & -25.429119 &938.969 &1272.19 &15.328 &0.813 \\
3-14	 & 117.595253 & -25.456421 &1611.64 &1345.44 &15.630 &0.634 \\
3-15	 & 117.599487 & -25.435949 &1107.33 &1440.01 &15.081 &0.781 \\
3-16*(4-17)	 & 117.603507 & -25.454220 &1557.44 &1529.28 &15.872 &0.420 \\
3-17*(4-18)	 & 117.605939 & -25.449539 &1442.39 &1583.72 &15.680 &0.369 \\
3-18	 & 117.610395 & -25.429110 &938.880 &1683.15 &15.278 &0.674 \\
3-19*(4-20)	 & 117.613478 & -25.439489 &1194.70 &1751.76 &15.960 &0.712 \\
3-20	 & 117.624078 & -25.442789 &1276.21 &1987.28 &15.261 &0.661 \\
3-21	 & 117.629099 & -25.429670 &953.049 &2099.37 &15.736 &0.497 \\
3-22*(4-23)	 & 117.633748 & -25.483009 &2267.17 &2202.25 &15.491 &0.444 \\
4-01*(3-01)	 & 117.544506 & -25.468800 &1915.97 &216.179 &15.472 &0.658 \\
4-02	 & 117.548947 & -25.436581 &1122.16 &315.220 &15.813 &0.828 \\
4-03	 & 117.553067 & -25.461161 &1727.81 &406.920 &15.976 &0.581 \\
4-04*(3-04)	 & 117.556415 & -25.449940 &1451.59 &481.209 &15.420 &0.927 \\
4-05	 & 117.558961 & -25.462400 &1758.43 &537.789 &16.026 &0.779 \\
4-06*(3-06)	 & 117.561121 & -25.449490 &1440.56 &585.820 &15.961 &0.702 \\
4-07*(3-07)	 & 117.563975 & -25.453800 &1546.66 &649.270 &15.734 &0.476 \\
4-08	 & 117.566943 & -25.488420 &2399.71 &715.469 &15.274 &0.542 \\
4-09*(3-09)	 & 117.570255 & -25.452900 &1524.51 &789.250 &16.015 &0.542 \\
4-10	 & 117.577264 & -25.457500 &1637.97 &945.270 &15.956 &0.679 \\
4-11a	 & 117.581971 & -25.452629 &1518.19 &1049.93 &15.532 &0.868 \\
4-11b	 & 117.580473 &	-25.452248 &1509.52 &1016.56&15.571 &0.596 \\
4-12	 & 117.585704 & -25.469730 &1939.42 &1133.10 &15.383 &0.361 \\
4-13a	 & 117.588186 & -25.474010 &2045.04 &1188.56 &15.475 &0.935 \\
4-13b	 & 117.588623 &	-25.474132 &2048.10 &1198.28&15.695 &0.708 \\
4-14a	 & 117.590911 & -25.451870 &1499.50 &1249.11 &15.888 & 0.423 \\
4-14b	 & 117.591291 &	-25.451637 &1493.77	&1257.58&16.310 &0.603 \\
4-15	 & 117.593422 & -25.431339 &993.710 &1305.04 &15.612 &0.877 \\
4-16	 & 117.600982 & -25.482161 &2245.90 &1472.66 &15.475 &0.560 \\
4-17*(3-16)	 & 117.603507 & -25.454220 &1557.44 &1529.28 &15.872 &0.420 \\
4-18*(3-17)	 & 117.605939 & -25.449539 &1442.39 &1583.72 &15.680 &0.368 \\
4-19*(5-15,6-07)	 & 117.611318 & -25.436760 &1127.38 &1703.61 &15.817 &0.504 \\
4-20*(3-19)	 & 117.613478 & -25.439489 &1194.70 &1751.76 &15.960 &0.712 \\
4-21a	 & 117.624779 & -25.485121 &2319.12 &2002.35 &15.255 &0.439 \\
4-21b	 & 117.625425 &	-25.484931 &2314.45	&2016.80&16.869 &0.370 \\
4-22*(5-19,6-10)	 & 117.629156 & -25.445040 &1331.68 &2100.31 &15.105 &0.656 \\
4-23*(3-22)	 & 117.633748 & -25.483009 &2267.17 &2202.25 &15.491 &0.444 \\
5-01*(6-01)	 & 117.633748 & -25.483009 &2507.56 &243.380 &15.376 &0.866 \\
5-02	 & 117.552874 & -25.432360 &1018.42 &402.420 &18.737 &1.728 \\
5-03	 & 117.554998 & -25.450260 &1459.31 &449.829 &18.998 &1.888 \\
5-04	 & 117.562079 & -25.495340 &2570.23 &607.570 &18.561 &1.613 \\
5-05	 & 117.566049 & -25.470680 &1962.70 &695.669 &18.758 &1.834 \\
5-06	 & 117.569346 & -25.461491 &1736.31 &769.010 &19.029 &1.846 \\
5-07	 & 117.572458 & -25.473909 &2042.33 &838.390 &18.861 &2.043 \\
5-08	 & 117.576156 & -25.466030 &1848.18 &920.820 &18.486 &1.642 \\
5-09a	 & 117.578151 & -25.468010 &1897.08 &964.890 &18.632 &1.755 \\
5-09b	 & 117.578932 &	-25.467887 &1894.04	&982.310&19.774&2.602 \\
5-10	 & 117.583730 & -25.453449 &1538.35 &1089.32 &18.375 &1.603 \\
5-11	 & 117.589753 & -25.429340 &944.349 &1223.56 &18.611 &1.682 \\
5-12	 & 117.598350 & -25.457541 &1639.22 &1414.72 &18.609 &1.613 \\
5-13	 & 117.601511 & -25.444740 &1324.00 &1484.90 &18.926 &1.960 \\
5-14a	 & 117.604973 & -25.433540 &1048.05 &1562.03 &18.905 &2.036 \\
5-14b	 & 117.604973 &	-25.433550 &1048.05	&1562.03 &         &  \\
5-15*(4-19,6-07)	 & 117.611318 & -25.436760 &1127.38 &1703.61 &15.817 &0.504 \\
5-16	 & 117.618692 & -25.435690 &1101.15 &1867.42 &18.906 &1.955 \\
5-17	 & 117.623498 & -25.471460 &1982.58 &1974.27 &18.763 &1.645 \\
5-18	 & 117.625844 & -25.438560 &1172.06 &2026.73 &18.299 &1.701 \\
5-19*(4-22,6-10)	 & 117.629156 & -25.445040 &1331.68 &2100.31 &15.105 &0.656 \\
5-20	 & 117.632053 & -25.462080 &1751.65 &2164.44 &18.477 &1.646 \\
5-21	 & 117.634120 & -25.435869 &1105.85 &2210.98 &18.641 &1.939 \\
6-01*(5-01)	 & 117.545714 & -25.492781 &2507.56 &243.380 &15.367 &0.866 \\
6-02a	 & 117.568680 &	-25.491604 &2478.14	&754.240&17.439&0.934 \\
6-02b	 & 117.569010 & -25.491421 &2473.57 &761.570 &18.834 &1.882 \\
6-03	 & 117.578845 & -25.468510 &1909.33 &980.159 &18.830 &1.757 \\
6-04a	 & 117.588501 &	-25.442119 &1259.17	&1195.62&18.691 &1.725 \\
6-04b	 & 117.589116 & -25.441950 &1255.00 &1209.34 &18.603 &1.794 \\
6-05	 & 117.596862 & -25.465771 &1842.04 &1381.34 &18.882 &2.064 \\
6-06	 & 117.604158 & -25.455910 &1599.11 &1543.75 &18.957 &1.981 \\
6-07*(4-19,5-15)	 & 117.611318 & -25.436760 &1127.38 &1703.61 &15.817 &0.504 \\
6-08	 & 117.616796 & -25.488850 &2410.82 &1824.96 &19.058 &1.870 \\
6-09	 & 117.623491 & -25.482040 &2243.32 &1973.60 &18.654 &1.670 \\
6-10*(4-22,5-19)	 & 117.629156 & -25.445040 &1331.68 &2100.31 &15.105 &0.656 \\
\tableline
\enddata
\end{deluxetable*}

\end{document}